\documentclass[10pt]{article}
\usepackage{makeidx}
\usepackage{amsmath}
\usepackage{amsfonts}
\usepackage{amssymb}
\usepackage{graphicx}
\usepackage{cite}
\usepackage{xcolor}
\usepackage{amsthm}

\makeatletter
\g@addto@macro\th@plain{\thm@headpunct{:}}
\makeatother
\theoremstyle{plain}
\newtheorem*{acknowledgement*}{Acknowledgement}
\input epsf.sty
\newtheorem{theorem}{Theorem}
\newtheorem{acknowledgement}[theorem]{Acknowledgement}

\makeatletter \@addtoreset{equation}{section}
\renewcommand{\theequation}{\thesection.\arabic{equation}}
\input{tcilatex}
\begin{document}

\title{\textbf{A signature index}\\
\textbf{\ for third order topological insulators}}
\author{L.B Drissi$^{1,2,3}$ and E.H Saidi$^{1,2}$ \\
{\small 1- LPHE, Modeling \& Simulations, Faculty of Science, Mohammed V
University, Rabat, Morocco.}\\
{\small 2- CPM, Centre of Physics and Mathematics, Mohammed V University in
Rabat, Morocco.}\\
{\small 3-\ Peter Gr\"{u}nberg Institut and Institute for Advanced
Simulation, }\\
{\small Forschungszentrum J\"{u}lich \& JARA, D-52425 J\"{u}lich, Germany. }}
\maketitle

\begin{abstract}
In this work, we develop an index signature characterising the third order
topological phases in 3D systems. This index is an alternating sum of
monomial signatures of Higgs triplet values at 3D corners. We extend our
method to N-dimensional systems with open boundaries, and demonstrate that
the topological invariant can be efficiently generalised to any space
dimension including the second order topological insulators. Known results
on lower dimensional systems are recovered and an interpretation in the
Higgs space parameters is given.\newline

\textbf{Keys words}: BBH lattice models; Second and Third Order topological
Insulators; Index theorem; Higgs space.
\end{abstract}

\section{Introduction}

Recently, a new family of topological insulators (HOTIs) going beyond the
standard \textrm{Altland-Zirnbauer} (AZ) classification \cite{1A,2A} has
been discovered by Benalcazar- Bernevig- Hughes (BBH) \cite{1B,2B}; see also
\cite{3B,5B} for related works. Higher Order Topological Insulators
---HOTIs--- essentially concern the set of 2D and 3D matter systems
exhibiting gapless modes at sub-regions of dimensions less or equal to $D-2$%
. In 2D systems, with polygonal shapes including the square and the
rectangle to be revisited in this study, \textrm{S}econd \textrm{O}rder
\textrm{T}opological \textrm{I}nsulators (SOTIs) are characterized by
gapless states existing only at corners while the surface and the edges are
gapped\textrm{\ }\cite{1C,2C,3C,4C,5C,6C,7C}\textrm{. }For 3D matter
however; one distinguishes two kinds of HOTIs, namely (i) a second
topological order phase on the 1-dimensional edges associated with a
periodic boundary condition in z-direction while x- and y- dimensions are
open\textrm{\ }\cite{3B}; and (ii)\textrm{\ }TOTIs; a \textrm{T}hird \textrm{%
O}rder \textrm{T}opological \textrm{I}nsulator for the point vertices
associated with full open boundaries\textrm{\ }\cite{1D}\textrm{; }this
third order topological phase concerns 3D poly-faces hosting gapless states
at corners (intersections of three 2D faces); while the bulk and the
boundary surface as well as their edge intersections are all of them gapped.
So far, TOTIs were constructed on the breathing Pyrochlore lattice where
each corner of the tetrahedron carries 1/4 fractional charge \textrm{\cite%
{3D}}. Hall conductance quantized in units of $e^{2}/h$\ was reported for
reflection symmetric second-order topological crystalline insulators where
the existence of edge states is ensured as long as surface and bulk remain
gapped\textrm{\  \cite{4B}. }Gapless corner states were observed
experimentally in a two-dimensional quadrupole topological insulator
implemented using perturbative mechanical metamaterials\textrm{\  \cite{5D}.
It was reported in \cite{5AD} that a realization of TOTIs was explored in an
anisotropic diamond- lattice acoustic metamaterial; the bulk acoustic band
structure has nontrivial topology characterized by quantized Wannier
centers. By acoustic measurement, gapless states were observed at two
corners of a rhombohedron- like structure in agreement with non trivial
topology characterized by quantized Wannier centers \cite{5BD}.} \textrm{%
\newline
}Topological invariants are so important for a more comprehensive
classification of topological crystalline insulators\textrm{\  \cite{2B,6D}. }%
Remarkably, this notion encodes information on the boundary physics and
provides access to natural quantities and observables\textrm{\  \cite%
{7D,8D,9D}. A noninteracting }$%
\mathbb{Z}
_{2}$\textrm{\ }topological invariant was given in term of the Berry
curvature for topological insulators (TIs)\textrm{\  \cite{10D}. }For
interacting and disordered TI systems, topological index, that determines
their phase diagrams, can be experimentally measured through the topological
magneto-electric effect\textrm{\  \cite{11D,12D,AZE3}. }For suitable types of
noise, the classification of mixed-state topology in one dimension reveals
retainment of its topological properties\textrm{\  \cite{13D}. }Furthermore,
topological Thouless pump is induced by Markovian reservoirs in open quantum
chains\textrm{\  \cite{14D}. }The integer invariant describing the topology
of 2D open systems captures the number difference of gapless edge modes and
gapless edge blind bands\textrm{\  \cite{15D}. }Remarkably, in the limit of
TIs, topological invariants are well investigated, however, they are not
perfectly developed and are to be described for HOTIs.\newline
In this paper, we derive an explicit formula for the topological invariant
characterising TOTIs in 3D parallelepiped systems with the cube as a
particular case. For that, we consider the topological DBI class of the AZ
periodic table which has reflection symmetries, in addition to the usual T-
P- C invariance. These kind of systems have full open boundary conditions
with topological dynamics remarkably described by the limits of the lattice
Hamiltonian near the Dirac points. It happens that the resulting limits can
be interpreted in terms of couplings between fermions and scalar fields as
done in \cite{1F}; and to which we refer below to as Fukui- coupling. This
nice observation has in fact a deep origin since from the quantum field
theory view (QFT), the coupling can be put in \emph{a formal correspondence}%
\footnote{%
\  \textrm{\ This correspondence is purely formal; it is used here to
emphasize the role played by the scalar field }$\mathbf{\phi }$\textrm{\ in
our calculations. Clearly this }$\mathbf{\phi }$\textrm{\ is not the true
Higgs scalar field of the standard model of elementary particles. There, the
Higgs field is a complex field doublet; it has a scalar potential }$\mathcal{%
V}\left( h\right) =-\mu ^{2}\left \vert h\right \vert ^{2}+\lambda \left
\vert h\right \vert ^{4}$\textrm{\ having a non trivial minimum }$h_{\min
}\neq 0$. \textrm{Moreover, the }$h$\textrm{\ couples to several fields of
the model; in particular to fermions like }$\psi ^{+}h\psi $\textrm{. In our
present study, the }$\phi $\textrm{\ scales in same manner as }$h$\textrm{\
and has a quite similar tri-coupling }$\psi ^{+}\phi \psi $\textrm{. Here,
the }$\phi $\textrm{\ is handled as in} \cite{1F}.}\emph{\ }with the well
known Yukawa tri-coupling $\mathbf{\psi }^{\mathbf{\dagger }}\mathbf{\phi
\psi }$ giving masses to fermions through a non zero vacuum expectation
value $\left \langle \mathbf{\phi }\right \rangle $ of the Higgs field $%
\mathbf{\phi } $; the fields $\mathbf{\psi }$ and $\mathbf{\psi }^{\mathbf{%
\dagger }}$ describes the fermionic states. Borrowing this idea and applying
it to 3D lattice systems, we develop the topological picture for
3-dimensional BBH system and show amongst others that the underlying
topological index $Ind(H_{3D})$ is given by%
\begin{equation}
Ind(H_{3D})=\dsum \limits_{p,q,s=\pm }\frac{pqs}{8}sgn\left(
A_{p}B_{q}C_{s}\right)  \label{A}
\end{equation}%
with $A_{\pm },B_{\pm },C_{\pm }$ are non vanishing constants whose meaning
may be imagined in terms of vacuum expectation values of an O$\left(
3\right) $ scalar (Higgs ) field triplet $\left( \phi _{x},\phi _{y},\phi
_{z}\right) $. These constants are the values of Higgs field at space
infinities; they will be discussed in details in the heart of the paper. We
also show that the above index formula has a remarkable factorisation as in
Eq.(\ref{3ih}) showing in turns that $Ind(H_{3D})$ is just an element of a
sequence with a generic term as follows%
\begin{equation}
Ind(H_{ND})=\dprod \limits_{i=1}^{N}\left[ \sum_{p_{i}=\pm }\frac{p_{i}}{2}%
sgn\left( A_{p_{i}}\right) \right]  \label{I}
\end{equation}%
By setting $N=3$, we recover exactly the index formula (\ref{A}); for $N=1$,
one obtains an index formula for the 1-dimensional Su- Schrieffer- Heeger
models (SSH) model \cite{1E} which reads here as $\frac{1}{2}%
(sgnA_{+}-sgnA_{-})$; and by setting $N=2$, we discover the Fukui formula
for $Ind(H_{2D})$ describing the second order topological systems of 2D
matter systems. This 2D index formula can be expressed like $\sum_{p,q=\pm }%
\frac{pq}{2}sgn\left( A_{p}B_{q}\right) $; it is also factorable like
\begin{equation}
Ind(H_{2D})=\frac{1}{2}\left[ sgn\left( A_{+}\right) -sgn\left( A_{-}\right) %
\right] \times \frac{1}{2}\left[ sgn\left( B_{+}\right) -sgn\left(
B_{-}\right) \right]
\end{equation}%
and remarkably descends for the 3-dimensional index formula $Ind(H_{3D})$ by
fixing the $C_{\pm }$ signatures of the third component $\phi _{z}$ of the
Higgs field triplet as $sgn\left( C_{+}\right) =+1$ and $sgn\left(
C_{-}\right) =-1$. \newline
The organisation of this paper is as follows: In section II, we revisit some
useful aspects of the 2D model with full open boundary conditions and
compute the topological index $Ind(H_{2D})$ by using two different methods
that will be commented at proper places; the one used by Fukui and an
equivalent one using the power of differential forms. In section III, we
shed more light on the derivation of the Fukui formula by using a direct
approach based on topological mappings. In section IV, we develop the
construction to 3- dimensions and show that the topological index $%
Ind(H_{3D})$ for the third order topological phase in DBI class with
reflection symmetries is given by Eq(\ref{A}). Section V is devoted to
conclusion and comments. \textrm{In the appendices A and B, we report some
technical details which have been omitted from the core of the paper in
order to keep the chain of ideas forward to the index derivation.}

\section{Two dimensional BBH model revisited}

Following \cite{1F}, the index of the Hamiltonian ---Ind(H)--- of the two
dimensional BBH lattice model with open boundary conditions can be
determined by studying the properties of a $2+2$ component fermion $\mathbf{%
\psi }=\left( \mathbf{\lambda },\mathbf{\chi }\right) $ near the four
Dirac-like points $\mathbf{k}_{\mathbf{n}\pi }^{\ast }$ of the model which
are equal to $\left( n_{x}\pi ,n_{y}\pi \right) $ with $n_{i}=0,1$; see
details given after Eq(\ref{01}) and further details in appendix A. Around
each one of these $\mathbf{k}_{\mathbf{n}\pi }^{\ast }$ points, the fermion $%
\mathbf{\psi }$ is coupled to an external 2D vector $\mathbf{\varphi }_{%
\mathbf{n}\pi }$ with constant components $(\varphi _{n_{x}\pi }^{x},\varphi
_{n_{y}\pi }^{y})$, thus playing the role of a mass --- i.e: a gap energy
between valence and conducting bands---. We refer below to this $\mathbf{%
\varphi }_{\mathbf{n}\pi }$ as constant $O\left( 2\right) $ Higgs field
doublet; its two components $\varphi _{n_{a}\pi }^{a}$ depend on the Dirac
point on which one rests; they are given by $\Delta _{a}+\cos \left(
n_{a}\pi \right) $ where $\Delta _{x}$ and $\Delta _{y}$ are hopping
parameters of the model. For example, near the Dirac point $n_{x}=n_{y}=0$,
the Higgs components $\varphi _{0}^{x}$ and $\varphi _{0}^{y}$ read
respectively like $\Delta _{x}+1$ and $\Delta _{y}+1$; while near another
point, say $n_{x}=0,n_{y}=\pi ,$ we have $(\varphi _{0}^{x},\varphi _{\pi
}^{y})$ with y-component like $\Delta _{y}-1$. So, given a Dirac point, the
interacting dynamics between fermion $\mathbf{\psi }$, its adjoint $\mathbf{%
\psi }^{\dagger }$ and the Higgs field is described by the following typical
Hamiltonian matrix
\begin{equation}
H=\mathbf{\Lambda }^{x}k_{x}+\mathbf{\Lambda }^{y}k_{y}+\mathbf{\Omega }%
^{x}\varphi _{x}+\mathbf{\Omega }^{y}\varphi _{y}  \label{01}
\end{equation}%
where the hermitian 4$\times $4 matrices $\mathbf{\Lambda }^{i}$\ and $%
\mathbf{\Omega }^{i}$\ and their basic properties will be specified below.
To fix ideas, the term\textrm{\ }$\mathbf{\Lambda }^{x}k_{x}+\mathbf{\Lambda
}^{y}k_{y}$ may be viewed as the leading contribution coming from the
expansion of $\mathbf{\Lambda }^{x}\sin k_{x}+\mathbf{\Lambda }^{y}\sin
k_{y} $ near the Dirac point\textrm{\ }$\left( k_{x},k_{y}\right) =\left(
0,0\right) $\textrm{\ }while the terms $\mathbf{\Omega }^{x}\varphi _{x}$
and $\mathbf{\Omega }^{y}\varphi _{y}$ derive from the expansion of\textrm{\
}$\mathbf{\Omega }^{x}\left( \Delta _{x}+\cos k_{x}\right) $ and $\mathbf{%
\Omega }^{y}\left( \Delta _{y}+\cos k_{y}\right) $ respectively \cite{1B,2B}%
. In appendix A, we show that the above H is in fact a representative of a
set of four Hamiltonians H$_{n_{x}\pi ,n_{y}\pi }$\ obtained by the
expansions of an underlying lattice H$_{lat}$\ near the points $\left(
k_{x},k_{y}\right) =\left( n_{x}\pi ,n_{y}\pi \right) $ with $%
n_{x},n_{y}=0,1 $ mod 2. Notice by the way that the denomination of the
usual four matrices $\mathbf{\gamma }^{\mu }$\ of Dirac theory by the pairs $%
\mathbf{\Lambda }^{i} $\ and $\mathbf{\Omega }^{i}$\ is for convenience; it
will be justified later on. Observe also that, in addition to the 2D
momentum vector variables $\left( k_{x},k_{y}\right) $, Eq(\ref{01}) depends
also on the constant $(\varphi _{x},\varphi _{y})$ that can be interpreted
at this level as moduli parameters and whose role they play will be explored
during this study. Notice as well that H has discrete symmetries given by
the basic $\boldsymbol{T}$- $\boldsymbol{P}$- $\boldsymbol{C}$ invariance%
\emph{\ }of AZ classification\  \cite{1A,2A} and moreover by mirror
symmetries going beyond AZ.\textrm{\ }The basic symmetries, respectively
generated by time reversing symmetry (TRS) operator $\boldsymbol{T}$,
particle-hole $\boldsymbol{P}$ and chiral operator $\boldsymbol{C}$, act on
H as follows%
\begin{eqnarray}
\boldsymbol{T}H\left( k_{x},k_{y}\right) \boldsymbol{T}^{-1} &=&+H\left(
-k_{x},-k_{y}\right)  \notag \\
\boldsymbol{P}H\left( k_{x},k_{y}\right) \boldsymbol{P}^{-1} &=&-H\left(
-k_{x},-k_{y}\right)  \label{c} \\
\boldsymbol{C}H\left( k_{x},k_{y}\right) \boldsymbol{C}^{-1} &=&-H\left(
k_{x},k_{y}\right)  \notag
\end{eqnarray}%
The obey the characteristics of the DBI class and are realised like\textrm{:
}$\boldsymbol{T}=K$\textrm{\ }with $K$ standing for the usual $\left( \ast
\right) $ complex conjugation; $\boldsymbol{C}=\mathbf{\gamma }_{5}$\textrm{%
\ }for chirality\textrm{\ }and\textrm{\ }$\boldsymbol{P}$ given by the
composed operator\textrm{\ }$\mathbf{\gamma }_{5}K.$\textrm{\ }Regarding the
mirror symmetries, they are generated by the $\boldsymbol{M}_{x}$ and $%
\boldsymbol{M}_{y}$ reflections in the x- and y- directions operating as
follows
\begin{eqnarray}
\boldsymbol{M}_{x}H\left( k_{x},k_{y}\right) \boldsymbol{M}_{x}^{-1}
&=&H\left( -k_{x},k_{y}\right)  \notag \\
\boldsymbol{M}_{y}H\left( k_{x},k_{y}\right) \boldsymbol{M}_{y}^{-1}
&=&H\left( k_{x},-k_{y}\right)  \label{d} \\
\boldsymbol{M}_{xy}H\left( k_{x},k_{y}\right) \boldsymbol{M}_{xy}^{-1}
&=&H\left( -k_{x},-k_{y}\right)  \notag
\end{eqnarray}%
with $\boldsymbol{M}_{xy}=\boldsymbol{M}_{y}\boldsymbol{M}_{x}$ for
inversion. The two $\boldsymbol{M}_{j}$ operators are realised by $i\mathbf{%
\Lambda }_{j}\mathbf{\gamma }_{5}$ and capture important properties \cite{4B}%
; they commute with $\boldsymbol{T}$ but anticommutes with\textrm{\ }$%
\mathbf{\gamma }_{5}$; the last property prevents a $\mathbf{\gamma }_{5}$
term in $H$. So, by requiring these reflection symmetries, the above time
reversing invariant Dirac Hamiltonian H has inevitably chiral symmetry which
is at the basis of our following calculations. Recall that the gap energy of
(\ref{01}) is induced by the constant $\mathbf{\varphi }$; gapless states
require then the vanishing of $\mathbf{\varphi }$. In the remainder of the
paper, we think of the topological nature of the corner states of same type
as the Jackiw- Rossi states \cite{JR}; for that we promote the constant $%
\varphi ^{a}$ to a smooth space coordinate dependent $\phi ^{a}=\phi
^{a}\left( x,y\right) $ with topological transition encoded by a sign change
of the Higgs field $\phi ^{a}$. In what follows, we substitute (\ref{01}) by
the relaxed Hamiltonian $\mathbf{\Lambda }^{i}k_{i}+\mathbf{\Omega }^{a}\phi
_{a}$ where $\phi ^{a}$ is coordinate dependent and summation over repeated
indices. \textrm{Regarding materials, notice that our present work} \textrm{%
concerns solids with corners and boundary edges related by reflection
symmetries; and so the results to be derived below apply to those 2D- and
3D- materials having discrete symmetry groups }$\mathcal{G}_{2D}$\textrm{\
and }$\mathcal{G}_{3D}$\textrm{\ containing reflections as sub-symmetries.}

\subsection{Integral formula for Ind(H)}

To characterise the topological properties of (\ref{01}) captured by gapless
states at corners of the rectangular system, we use chiral symmetry $%
\boldsymbol{C}$ of the model to calculate the index of the hamiltonian; this
is achieved by applying the method of \textrm{\cite{1F} }relying on the
fermion-Higgs coupling to reach Ind(H). For this purpose, let us begin by
giving some useful tools to fix ideas and which are also helpful when
studying the 3D system. The four energy bands of H are described by $2+2$
component fermion $\mathbf{\psi =\psi }\left( x,y\right) $ that we split like%
\begin{equation}
\mathbf{\psi }=\left(
\begin{array}{c}
\mathbf{\lambda } \\
\mathbf{\bar{\chi}}%
\end{array}%
\right)
\end{equation}%
\textrm{with} $\mathbf{\bar{\chi}}=\mathbf{\chi }^{\ast }$ \textrm{and where}
\begin{equation}
\mathbf{\lambda }=\left(
\begin{array}{c}
\lambda _{1} \\
\lambda _{2}%
\end{array}%
\right) \qquad ,\qquad \mathbf{\chi }=\left(
\begin{array}{c}
\chi _{1} \\
\chi _{2}%
\end{array}%
\right)
\end{equation}%
The explicit form of the wave functions describing gapless states at each
corner are obtained by solving the two following constraint relations:

\begin{enumerate}
\item The eigenvalue equation $H\mathbf{\psi }=E\mathbf{\psi }$ with energy $%
E=0$ and local hamiltonian as
\begin{equation}
H=-i\mathbf{\Lambda }^{j}\frac{\partial }{\partial x^{j}}+\mathbf{\Omega }%
^{a}\phi _{a}  \label{02}
\end{equation}%
where we have used $k_{j}=-i\frac{\partial }{\partial x^{j}}$. Here, the
matrices $\mathbf{\Lambda }^{j}$ and $\mathbf{\Omega }^{a}$ are realised in
terms of two sets of Pauli matrices $\mathbf{\sigma }$ and $\mathbf{\tau }$
as follows%
\begin{equation}
\begin{tabular}{lllllll}
$\mathbf{\Lambda }_{1}$ & $=$ & $\tau _{2}\otimes \sigma _{1}$ & $\qquad
,\qquad $ & $\mathbf{\Lambda }_{2}$ & $=$ & $\tau _{2}\otimes \sigma _{3}$
\\
$\mathbf{\Omega }_{1}$ & $=$ & $\tau _{2}\otimes \sigma _{2}$ & $\qquad
,\qquad $ & $\mathbf{\Omega }_{2}$ & $=$ & $\tau _{1}\otimes \sigma _{0}$%
\end{tabular}
\label{ma}
\end{equation}%
with the properties $\boldsymbol{T\mathbf{\Lambda }}_{i}\boldsymbol{T}^{-1}=-%
\boldsymbol{\mathbf{\Lambda }}_{i}$ and $\boldsymbol{T\mathbf{\Omega }}_{a}%
\boldsymbol{T}^{-1}=\mathbf{\Omega }_{a}$ while $\boldsymbol{C\boldsymbol{%
\mathbf{\Lambda }}_{i}C}^{-1}=-\boldsymbol{\mathbf{\Lambda }}_{i}$ and $%
\boldsymbol{C\boldsymbol{\mathbf{\Omega }}_{a}C}^{-1}=-\boldsymbol{\mathbf{%
\Omega }}_{a}$. The $\boldsymbol{\mathbf{\Lambda }}_{i}$ and $\boldsymbol{%
\mathbf{\Omega }}_{a}$ matrices square to identity and anticommute among
themselves; for a generic description of these matrices; it is interesting
to set $\mathbf{\Lambda }_{1}=\mathbf{\gamma }_{1},\mathbf{\Omega }_{1}=%
\mathbf{\gamma }_{2}$ and $\mathbf{\Lambda }_{2}=\mathbf{\gamma }_{3},%
\mathbf{\Omega }_{2}=\mathbf{\gamma }_{4}$; by using these $\mathbf{\gamma }%
_{\mu }$'s, the underlying 4-dim Euclidian Clifford algebra reads as $%
\mathbf{\gamma }_{\mu }\mathbf{\gamma }_{\nu }+\mathbf{\gamma }_{\nu }%
\mathbf{\gamma }_{\mu }=2\delta _{\mu \nu }$ and the chiral operator is
given by $\mathbf{\gamma }_{5}=-\mathbf{\gamma }_{1}\mathbf{\gamma }_{2}%
\mathbf{\gamma }_{3}\mathbf{\gamma }_{4}$; it represents the $\boldsymbol{C}$
in Eq.(\ref{c}) and defines chirality of the states. As given below, chiral
and antichiral fermions have the half degrees of freedom carried by $\mathbf{%
\psi },$
\begin{equation}
\mathbf{\gamma }_{5}\left(
\begin{array}{c}
\mathbf{\lambda } \\
\mathbf{0}%
\end{array}%
\right) =-\left(
\begin{array}{c}
\mathbf{\lambda } \\
\mathbf{0}%
\end{array}%
\right) \qquad ,\qquad \mathbf{\gamma }_{5}\left(
\begin{array}{c}
\mathbf{0} \\
\mathbf{\bar{\chi}}%
\end{array}%
\right) =\left(
\begin{array}{c}
\mathbf{0} \\
\mathbf{\bar{\chi}}%
\end{array}%
\right)
\end{equation}

\item A chirality condition, killing the half of the components of $\mathbf{%
\psi }$, is given by the equation
\begin{equation}
\mathbf{\gamma }_{5}\mathbf{\psi }=q_{\mathbf{\gamma }}\mathbf{\psi }\qquad
,\qquad \mathbf{\gamma }_{5}=-\tau _{3}\otimes \sigma _{0}
\end{equation}%
The chiral charge $q_{\mathbf{\gamma }}$ takes either the value $+1$ or $-1$%
; the exact value of this charge at corner is fixed by the normalisation
condition of $\mathbf{\psi }\left( \mathbf{r}\right) $. By taking $\phi _{a}$
positive constants for instance, the $\mathbf{\psi }$ is given by a real
exponential $e^{-\mathbf{\kappa }.\mathbf{r}}$ with momentum $\kappa $ which
is function of $q_{\mathbf{\gamma }}$; by demanding $\mathbf{\kappa }.%
\mathbf{r}>0,$ one ends with a constraint on the sign of the chiral charge;
\textrm{see} \textrm{\cite{4F} for details}.
\end{enumerate}

\  \  \  \newline
After this digression on $\mathbf{\psi }$ and the algebra of the $\mathbf{%
\Lambda }^{j}$ and $\mathbf{\Omega }^{a}$ matrices, we come now to the study
of the index of Hamiltonian H; this index can be expressed in different, but
equivalent, ways. The standard definition is given by the integer
\begin{equation}
Ind\left( H\right) =N_{+}-N_{-}
\end{equation}%
with $N_{\pm }$ standing for the numbers of chiral and antichiral zero modes
in the ground state. To compute the value of this index, it is helpful to
take advantage of two interesting features: First, we think of this integer
as the flux of some two dimensional current $J^{i}=J^{i}\left( \mathbf{r}%
\right) $ as follows%
\begin{equation}
Ind\left( H\right) =\doint \nolimits_{\mathcal{C}}\varpi \qquad with\qquad
\varpi =\frac{1}{2}\left( J^{x}dy-J^{y}dx\right)  \label{int}
\end{equation}%
with closed contour $\mathcal{C}$ belonging to the x-y plane. Below, we
refer to this position plane as $\mathbb{R}_{\mathbf{r}}^{2}$ and so the
loop $\mathcal{C}$ sits in $\mathbb{R}_{\mathbf{r}}^{2}$; the reason for
this notation is that we will encounter below another closed curve $\Gamma $
belonging to another plane $\mathbb{R}_{\mathbf{\phi }}^{2}$ parameterised
by the $\mathbf{\phi }=(\phi _{x},\phi _{y})$ components of the Higgs field
doublet. By using the antisymmetric Levi-Civita symbol $\varepsilon ^{ij}$%
\textrm{\ and its inverse }$\varepsilon _{ji}$ \textrm{with} $\varepsilon
^{12}=\varepsilon _{21}=1$, the above index reads in a generic form like
\begin{equation}
Ind\left( H\right) =-\frac{1}{2}\doint \nolimits_{\mathcal{C}}J^{i}\left(
\varepsilon _{ij}dx^{j}\right) .  \label{in}
\end{equation}%
showing that Ind(H) is indeed the flux of the $J_{i}$ vector. Second, we
interpret the vector current $J^{i}$ as the vacuum expectation value (VEV)
of an axial vector current $J_{5}^{i}$ involving the matrix operator $%
\mathbf{\gamma }_{5}$ generating chirality. In other words, we have $%
J^{i}=\left \langle J_{5}^{i}\right \rangle $ reading in terms of the
Hamiltonian H (\ref{01}-\ref{02}) as follows \textrm{\cite{1F,4F,1G,2G,3G,4G}%
},%
\begin{equation}
J^{i}\left( \mathbf{r}\right) =\lim_{\mathbf{r}^{\prime }\mathbf{\rightarrow
r}}\left[ \lim_{m\rightarrow 0}tr\left( \mathbf{\gamma }_{5}\mathbf{\Lambda }%
^{i}\frac{1}{m+iH}\right) \mathbf{\delta }_{2}\left( \mathbf{r}-\mathbf{r}%
^{\prime }\right) \right] .  \label{j}
\end{equation}%
By calculating the trace over the 4$\times $4 matrix product in (\ref{j})
and taking the vanishing mass limit, the above vector current can be
expressed in terms of the Higgs field as \textrm{follows}%
\begin{equation}
J^{i}=\frac{1}{\pi \left \vert \mathbf{\phi }\right \vert ^{2}}\varepsilon
^{ij}\phi _{a}\partial _{j}\phi _{b}\varepsilon ^{ab}.  \label{cu}
\end{equation}%
with $\left \vert \mathbf{\phi }\right \vert ^{2}=\phi _{x}^{2}+\phi
_{y}^{2} $. \textrm{The derivation of Eq.(\ref{cu}) is somehow technical; a
sketch of its obtention is reported in appendix B. }By substituting this
current vector back into (\ref{in}), we end up with the following integral
formula%
\begin{equation}
Ind\left( H\right) =\doint \nolimits_{\mathcal{C}}\frac{1}{2\pi \left \vert
\mathbf{\phi }\right \vert ^{2}}\varepsilon ^{ab}\phi _{a}\partial _{j}\phi
_{b}dx^{j}  \label{inf}
\end{equation}%
This relation shows that Ind$\left( H\right) $ is intimately related with
the Higgs fields; thank to fermion-Higgs coupling which allowed to extract
the ground state data $N_{+}-N_{-}$.

\subsection{Signature formula for Ind(H)}

\begin{figure}[tbph]
\begin{center}
\hspace{0cm} \includegraphics[width=8cm]{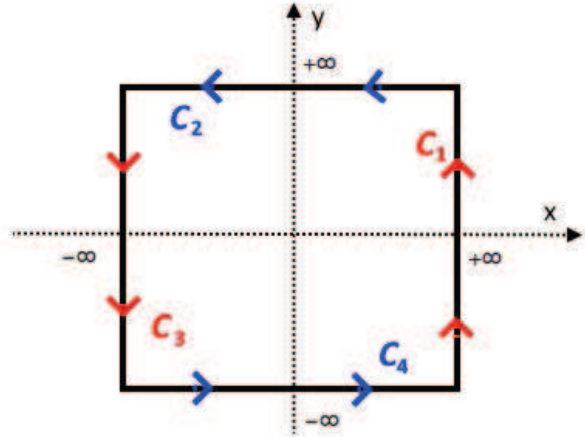}
\end{center}
\par
\vspace{-1 cm}
\caption{Flux of the vector current through the rectangular boundary at
space infinity \ This curve choice allows to collect the full flux.}
\label{c1}
\end{figure}

To calculate Ind(H), we must perform the integral in (\ref{inf}) which can
be interpreted as a function $\mathcal{F}\left[ \mathcal{C}\right] $ with
variable given by an extended object namely the loop $\mathcal{C}$. For
that, we a priori have to specify the closed curve $\mathcal{C}$ in order to
get the numerical value of Ind(H). In fact, this is not necessary since
Ind(H) is a topological invariant and so it is non sensitive to the shape of
$\mathcal{C}$; we will use this property later on in order to re-derive the
result of \textrm{\cite{1F}}. Before that, let us follow the \textrm{Fukui}
approach by thinking of the curve in Eq.(\ref{inf}) as a big closed $%
\mathcal{C}_{\infty }$ at the boundary of the x-y plane; and moreover
imagine it as a rectangular loop $\mathcal{C}_{\infty }^{\left( 4\right) }$
with four edges as depicted in the \textbf{Figure \ref{c1}}.
\begin{equation}
\mathcal{C}_{\infty }^{\left( 4\right) }=\mathcal{C}_{1}\cup \mathcal{C}%
_{2}\cup \mathcal{C}_{3}\cup \mathcal{C}_{4}  \label{ch}
\end{equation}%
We also require that at the four corners of the rectangular curve $\mathcal{C%
}_{\infty }^{\left( 4\right) }$, the Higgs components $\left( \phi _{x},\phi
_{y}\right) $ take constant values denoted below like $\left( A_{\pm
},B_{\pm }\right) $.
In other words, on the boundary edges of $\mathcal{C}_{\infty }^{\left(
4\right) }$, the Higgs field components satisfy space boundary conditions of
the type%
\begin{equation}
\begin{tabular}{lllll}
$\lim_{x\rightarrow \pm \infty }\phi _{x}\left( x,y\right) $ & $=$ & $\phi
_{x}\left( \pm \infty ,y\right) $ & $=$ & $A_{\pm }$ \\
$\lim_{x\rightarrow \pm \infty }\phi _{y}\left( x,y\right) $ & $=$ & $\phi
_{y}\left( \pm \infty ,y\right) $ & $=$ & $G\left( y\right) $%
\end{tabular}%
\end{equation}%
and
\begin{equation}
\begin{tabular}{lllll}
$\lim_{y\rightarrow \pm \infty }\phi _{y}\left( x,y\right) $ & $=$ & $\phi
_{y}\left( x,\pm \infty \right) $ & $=$ & $B_{\pm }$ \\
$\lim_{y\rightarrow \pm \infty }\phi _{x}\left( x,y\right) $ & $=$ & $\phi
_{x}\left( x,\pm \infty \right) $ & $=$ & $F\left( x\right) $%
\end{tabular}%
\end{equation}%
where $A_{\pm }$ and $B_{\pm }$ are non vanishing constants. We also have $%
\lim_{x\rightarrow \pm \infty }F\left( x\right) =A_{\pm }$ and $%
\lim_{y\rightarrow \pm \infty }G\left( y\right) =B_{\pm }$. This implies
that the value of the field components $\left( \phi _{x},\phi _{y}\right) $
at the corners of the rectangle are precisely given by $\left( A_{\pm
},B_{\pm }\right) $. With these boundary conditions the vector current (\ref%
{cu}) on the curve $\mathcal{C}_{\infty }$ becomes%
\begin{equation}
\begin{tabular}{lll}
$\left. J^{x}\left( x,y\right) \right \vert _{\mathcal{C}_{1}}dy$ & $=$ & $%
\frac{\varepsilon ^{ab}}{\pi \left( A_{+}^{2}+G^{2}\right) }G_{a}\partial
_{y}G_{b}dy$ \\
$\left. J^{y}\left( x,y\right) \right \vert _{\mathcal{C}_{2}}dx$ & $=$ & $%
\frac{-\varepsilon ^{ab}}{\pi \left( B_{+}^{2}+F^{2}\right) }F_{a}\partial
_{x}F_{b}dx$%
\end{tabular}%
\end{equation}%
that read also like
\begin{equation}
\begin{tabular}{lll}
$\left. J^{x}\left( x,y\right) \right \vert _{\mathcal{C}_{1}}dy$ & $=$ & $%
\frac{1}{\pi }\frac{dg}{1+g^{2}}$ \\
$\left. J^{y}\left( x,y\right) \right \vert _{\mathcal{C}_{2}}dx$ & $=$ & $%
\frac{1}{\pi }\frac{df}{1+f^{2}}$%
\end{tabular}%
\end{equation}%
where we have set $g=\frac{G}{A_{+}}$ and $f=\frac{F}{B_{+}}$. Because of
the property $\frac{d\xi }{1+\xi ^{2}}=d\left( \arctan \xi \right) $, the
above $\left. J_{x}\right \vert _{\mathcal{C}_{1}}dy$ and $\left.
J_{y}\right \vert _{\mathcal{C}_{2}}dx$ are exact 1-forms given by%
\begin{equation}
\begin{tabular}{lll}
$\left. J^{x}\left( x,y\right) \right \vert _{\mathcal{C}_{1}}dy$ & $=$ & $%
\frac{1}{\pi }d\left( \arctan g\right) $ \\
$\left. J^{y}\left( x,y\right) \right \vert _{\mathcal{C}_{2}}dx$ & $=$ & $%
\frac{1}{\pi }d\left( \arctan f\right) $%
\end{tabular}
\label{GF}
\end{equation}%
Similarly, we have%
\begin{equation}
\begin{tabular}{lll}
$\left. J^{x}\left( x,y\right) \right \vert _{\mathcal{C}_{3}}dy$ & $=$ & $%
\frac{1}{\pi }d(\arctan \tilde{g})$ \\
$\left. J^{y}\left( x,y\right) \right \vert _{\mathcal{C}_{4}}dx$ & $=$ & $%
\frac{1}{\pi }d(\arctan \tilde{f})$%
\end{tabular}%
\end{equation}%
with $\tilde{g}=\frac{\tilde{G}}{A_{-}}$ and $\tilde{f}=\frac{\tilde{F}}{%
B_{-}}$. Substituting these relations back into Eq. (\ref{int}), and using
\begin{equation}
\begin{tabular}{lll}
$\dint \nolimits_{\mathcal{C}_{1}}\varpi $ & $=$ & $\frac{1}{2\pi }\arctan
\frac{B_{+}}{A_{+}}-\frac{1}{2\pi }\arctan \frac{B_{-}}{A_{+}}$ \\
$\dint \nolimits_{\mathcal{C}_{2}}\varpi $ & $=$ & $\frac{1}{2\pi }\arctan
\frac{A_{+}}{B_{+}}-\frac{1}{2\pi }\arctan \frac{A_{-}}{B_{+}}$%
\end{tabular}
\label{c12}
\end{equation}%
as well as%
\begin{equation}
\begin{tabular}{lll}
$\dint \nolimits_{\mathcal{C}_{3}}\varpi $ & $=$ & $\frac{1}{2\pi }\arctan
\frac{B_{-}}{A_{-}}-\frac{1}{2\pi }\arctan \frac{B_{+}}{A_{-}}$ \\
$\dint \nolimits_{\mathcal{C}_{4}}\varpi $ & $=$ & $\frac{1}{2\pi }\arctan
\frac{A_{-}}{B_{-}}-\frac{1}{2\pi }\arctan \frac{A_{+}}{B_{-}}$%
\end{tabular}
\label{c34}
\end{equation}%
we end up with the following value of the index%
\begin{equation}
\begin{tabular}{lll}
$IndH$ & $=$ & $\frac{1}{2\pi }(\arctan \frac{B_{+}}{A_{+}}+\arctan \frac{%
A_{+}}{B_{+}})-\frac{1}{2\pi }(\arctan \frac{A_{-}}{B_{+}}+\arctan \frac{%
B_{+}}{A_{-}})$ \\
&  & $\frac{1}{2\pi }(\arctan \frac{B_{-}}{A_{-}}+\arctan \frac{A_{-}}{B_{-}}%
)-\frac{1}{2\pi }(\arctan \frac{A_{+}}{B_{-}}+\arctan \frac{B_{-}}{A_{+}})$%
\end{tabular}%
\end{equation}%
Using the relation $\arctan \xi +\func{arccot}\xi =\frac{\pi }{2}sgn\left(
\xi \right) $, we can put the above relation into the following form%
\begin{equation}
\begin{tabular}{lll}
$Ind\left( H\right) $ & $=$ & $+\frac{1}{4}\left[ sgn\left(
A_{+}B_{+}\right) +sgn\left( A_{-}B_{-}\right) \right] $ \\
&  & $-\frac{1}{4}\left[ sgn\left( A_{+}B_{-}\right) +sgn\left(
A_{-}B_{+}\right) \right] $%
\end{tabular}
\label{ind}
\end{equation}%
which shows that $Ind\left( H\right) $ takes indeed an integer value; this
is the Fukui formula \cite{1F}. For the example where $A_{\pm }$, $B_{\pm }$
have the same sign, the index vanishes identically; it vanishes also for
other cases like $sgn\left( A_{\pm }\right) =1$ and $sgn(B_{-})=-sgn\left(
B_{+}\right) $. However, for the cases $A_{+}$, $B_{+}$ positive (resp.
negative) definite numbers and $A_{-},B_{-}$ negative (resp. positive)
definite ones; the index of the Hamiltonian is equal to $+1$ (resp. $-1$).
For the example $A_{-}=-A_{+}$ and $B_{-}=-B_{+}$, the index of H reduces to
$sgn\left( A_{+}B_{+}\right) $ which can be either $+1$ or $-1$ depending on
the signs of $A_{+}$ and $B_{+}$; for instance this is the case of $%
A_{+}=B_{+}$ and $A_{+}=-B_{+}$.

\section{More on Ind(H) in 2-dim DBI}

In this section, we shed more light of the derivation of the Fukui formula
by using a direct approach based on topological mapping between the curve $%
\mathcal{C}$ in the x-y position space $\mathbb{R}_{\mathbf{r}}^{2}$ and a
corresponding curve $\Gamma $ in the $\phi _{x}$-$\phi _{y}$ Higgs space $%
\mathbb{R}_{\mathbf{\phi }}^{2}$. We show amongst others that for a non
trivial index the $sgn\left( \phi _{x}\right) $ and $sgn\left( \phi
_{y}\right) $ have to change their polarity under space reflections.

\subsection{Working in Higgs plane}

\begin{figure}[tbph]
\begin{center}
\hspace{0cm} \includegraphics[width=7cm]{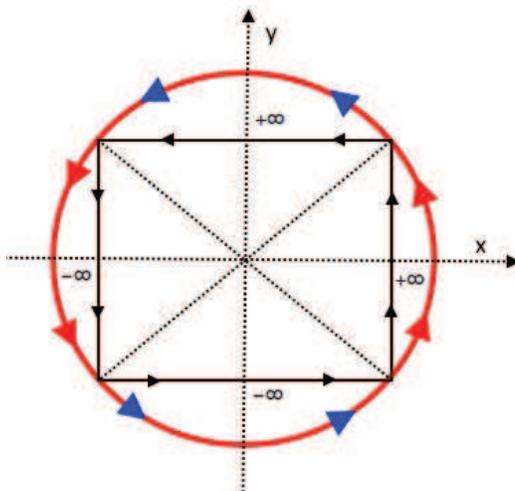}
\end{center}
\par
\vspace{-1 cm}
\caption{This figure has two interpretations depending on whether we are
sitting in $\mathbb{R}_{\mathbf{r}}^{2}$ or in $\mathbb{R}_{\mathbf{\protect%
\phi }}^{2}$. From the view of $\mathbb{R}_{\mathbf{r}}^{2}$, we have a
square $\mathcal{C}_{\infty }^{\left( 4\right) }$ in the x-y plane
circumscribed in a circle $\mathbb{S}_{\mathbf{r}}^{1}$ in red color. Under
the mapping $\Phi :\mathbf{r\rightarrow \protect \phi }\left( \mathbf{r}%
\right) $, this figure can be also interpreted in terms of closed curve $%
\Gamma _{\infty }^{\left( 4\right) }$ circumscribed in a circle $\mathbb{S}_{%
\mathbf{\protect \phi }}^{1}$ in the Higgs plane $\mathbb{R}_{\mathbf{\protect%
\phi }}^{2}$.}
\label{2}
\end{figure}

\textrm{The index formula }(\ref{ind}) obtained by using the curve choice (%
\ref{ch}) is remarkable and suggestive; it involves only $sgn\left( X\right)
$ functions in agreement with the topological aspect of the index; it is
given by the sum of four terms; each contributing with $\pm \frac{1}{4}$ and
add exactly to an integer. This relation shows that the four terms in (\ref%
{ind}) are in one to one correspondence with the four $\left( A_{\pm
},B_{\pm }\right) $ corners of a rectangular curve $\Gamma _{\infty
}^{\left( 4\right) }$ in the Higgs plane parameterised by $\left( \phi
_{x},\phi _{y}\right) $. For the case $A_{\pm }=B_{\pm }=\pm L$, the curve $%
\Gamma _{\infty }^{\left( 4\right) }$ reduces to a square in $\mathbb{R}_{%
\mathbf{\phi }}^{2}$, which can be imagined as depicted in the \textbf{%
Figure \ref{2}},
Below, we want to re-derive Eq. (\ref{ind}) by working directly in the Higgs
space $\mathbb{R}_{\mathbf{\phi }}^{2}$. This result, which relies on using
some differential geometry tools and topological mappings between $\mathbb{R}%
_{\mathbf{r}}^{2}$ and $\mathbb{R}_{\mathbf{\phi }}^{2}$, will be also used
later on when we study the third order topological index of the three
dimensional BBH lattice model. \newline
Substituting the vector current $J^{i}\left( \mathbf{r}\right) $ given by
Eq.(\ref{cu}) into the relation Eq.(\ref{in}) of the Hamiltonian index in
the x-y plane namely%
\begin{equation}
Ind\left( H\right) =\doint \nolimits_{\mathcal{C}}\varpi \qquad with\qquad
\varpi =-\frac{1}{2}\varepsilon _{ij}J^{i}dx^{j}
\end{equation}%
we obtain a new formula for the index which is completely expressed in the $%
\phi _{x}$-$\phi _{y}$ Higgs plane; it is given by%
\begin{equation}
Ind\left( H\right) =\doint \nolimits_{\Gamma }\frac{1}{2\pi \left \vert
\mathbf{\phi }\right \vert ^{2}}\varepsilon ^{ab}\phi _{a}d\phi _{b}
\label{iha}
\end{equation}%
where now the loop $\Gamma $ belongs to the $\phi _{x}$-$\phi _{y}$ Higgs
plane; it is the image of the closed curve $\mathcal{C}$ under the mapping $%
\mathbf{\phi }_{a}:\left( x,y\right) \rightarrow \left( \phi _{x},\phi
_{y}\right) $. The above equation shows that Ind(H) has a pole singularity
at $\left \vert \mathbf{\phi }\right \vert =0$ indicating that gapless
states live there. To deal with (\ref{iha}), which reads also like%
\begin{equation}
Ind\left( H\right) =\doint \nolimits_{\Gamma }\mathcal{B}_{a}dl^{a}\qquad
,\qquad dl^{a}=\varepsilon ^{ab}d\phi _{b}
\end{equation}%
with $\mathcal{B}_{a}=\frac{\phi _{a}}{2\pi \left \vert \mathbf{\phi }%
\right
\vert ^{2}}$, it is interesting to use new field variables $\varrho
=\varrho \left( x,y\right) $ and $\vartheta =\vartheta \left( x,y\right) $
related to the orders as%
\begin{equation}
\begin{tabular}{lllllll}
$\phi _{x}$ & $=$ & $\varrho \cos \vartheta $ & \qquad ,\qquad & $\varrho $
& $=$ & $\sqrt{\phi _{x}^{2}+\phi _{y}^{2}}$ \\
$\phi _{y}$ & $=$ & $\varrho \sin \vartheta $ & \qquad ,\qquad & $\vartheta $
& $=$ & $\arctan \frac{\phi _{y}}{\phi _{x}}$%
\end{tabular}
\label{wn}
\end{equation}%
This Higgs field change allows to bring the above index relation into the
simple form%
\begin{equation}
Ind\left( H\right) =\doint \nolimits_{\Gamma }\frac{d\vartheta }{2\pi }
\label{dh}
\end{equation}%
which is easy to deal with. To exhibit the pole singularity, one may also
use the complex variables $w=\phi _{x}+i\phi _{y}=\varrho e^{i\vartheta }$
and $\bar{w}=\phi _{x}-i\phi _{y}=\varrho e^{-i\vartheta }$ and replace $%
\frac{d\vartheta }{2\pi }$ either by $\frac{dw}{2\pi iw}$ or $-\frac{d\bar{w}%
}{2\pi i\bar{w}}$; one ends up with the typical relation $\frac{\pm 1}{2\pi i%
}\doint \nolimits_{\Gamma }\frac{dw}{w}$ having non zero values for contour $%
\Gamma $\ surrounding the pole at $w=0$.

\subsection{Ind(H) in Higgs plane}

Clearly, the explicit computation of the integral Eq.(\ref{dh}) depends a
priori on the shape of the closed curve $\Gamma $. If thinking of this loop $%
\Gamma $ as an oriented circle with perimeter 2$\pi $; we end up with $%
Ind\left( H\right) =\pm 1$ depending on the sense of orientation of the
circle; one may also get an integer $n$ for the case of an oriented circle
with a winding number\footnote{%
\ This number may be exhibited in our calculations by considering the
general parametrisation $\phi _{x}=\varrho \cos \left( n\vartheta \right) $
and $\phi _{y}=\varrho \sin \left( n\vartheta \right) $ instead of\ Eq(\ref%
{wn}) corresponding to setting $n=1$.} n where the perimeter of $\Gamma $\
is $2\pi n$. However, for a loop $\Gamma $\ given by a rectangular
---square--- cycle made of four edges as in the \textbf{Figures \ref{2}-\ref%
{3}};
\begin{figure}[tbph]
\begin{center}
\hspace{0cm} \includegraphics[width=7cm]{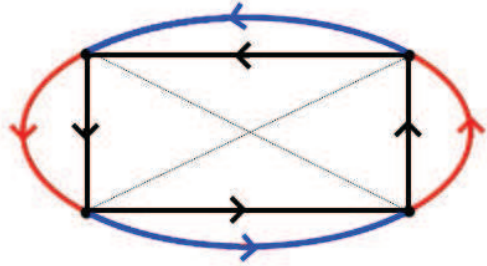}
\end{center}
\par
\vspace{-1 cm}
\caption{A rectangular loop $\Gamma _{\infty }^{\left( 4\right) }$
circumscribed into an ellipse in the Higgs plane. One can move continuously
from the ellipse to the rectangle by deforming the red and blue arcs of
circles into straight line segments.}
\label{3}
\end{figure}
one expects to obtain a relation like in (\ref{ind}). In this rectangular
case, we have%
\begin{equation}
\Gamma _{\infty }^{\left( 4\right) }=\Gamma _{1}^{+}\cup \Gamma _{2}^{+}\cup
\Gamma _{1}^{-}\cup \Gamma _{2}^{-}  \label{g12}
\end{equation}%
where the $\Gamma _{i}^{\pm }$'s are as%
\begin{equation}
\left.
\begin{tabular}{lllll}
$\Gamma _{1}^{\pm }$ & : & $B_{-}\leq \phi _{y}\leq B_{+}$ & ; & $\phi
_{x}=A_{\pm }$ \\
$\Gamma _{2}^{\pm }$ & : & $A_{-}\leq \phi _{x}\leq A_{+}$ & ; & $\phi
_{y}=B_{\pm }$%
\end{tabular}%
\right.
\end{equation}%
where, to fix the ideas, we have assumed $A_{-}<A_{+}$ and $B_{-}<B_{+}$;
but $A_{\pm }$ and $B_{\pm }$ can be either positive or negative showing
that one can distinguish several cases. From these $\Gamma _{i}^{\pm }$
sets, we learn that they can be also defined by using ratios like
\begin{equation}
\begin{tabular}{lll}
$\Gamma _{1}^{\pm }$ & $:$ & $\frac{B_{-}}{A_{\pm }}\leq \frac{\phi _{y}}{%
\phi _{x}}\leq \frac{B_{+}}{A_{\pm }}$ \\
$\Gamma _{2}^{\pm }$ & $:$ & $\frac{A_{-}}{B_{\pm }}\leq \frac{\phi _{x}}{%
\phi _{y}}\leq \frac{A_{+}}{B_{\pm }}$%
\end{tabular}%
\end{equation}%
teaching us that $\frac{\phi _{y}}{\phi _{x}}=\arctan \vartheta $ and its
inverse $\frac{\phi _{x}}{\phi _{y}}=\func{arccot}\vartheta $ are the
natural variables to use for describing the square loop $\Gamma _{\infty
}^{\left( 4\right) }$. Setting $\tan \vartheta _{pq}=\frac{B_{q}}{A_{p}}$
and $\cot \vartheta _{pq}=\frac{A_{p}}{B_{q}}$, we then have%
\begin{equation}
\begin{tabular}{lll}
$\Gamma _{1}^{\pm }$ & $:$ & $\arctan \frac{B_{-}}{A_{\pm }}\leq \vartheta
\leq \arctan \frac{B_{+}}{A_{\pm }}$ \\
$\Gamma _{2}^{\pm }$ & $:$ & $\func{arccot}\frac{A_{-}}{B_{\pm }}\leq
\vartheta \leq \func{arccot}\frac{A_{+}}{B_{\pm }}$%
\end{tabular}
\label{gpm}
\end{equation}%
Putting the decomposition (\ref{g12}) back into the index formula (\ref{dh})
and expanding, we obtain%
\begin{equation}
Ind\left( H\right) =\dint \nolimits_{\Gamma _{1}^{+}}\frac{d\vartheta }{2\pi
}+\dint \nolimits_{\Gamma _{2}^{+}}\frac{d\vartheta }{2\pi }+\dint
\nolimits_{\Gamma _{1}^{-}}\frac{d\vartheta }{2\pi }+\dint \nolimits_{\Gamma
_{2}^{-}}\frac{d\vartheta }{2\pi }
\end{equation}%
Then, using (\ref{gpm}), we end up with
\begin{equation}
\begin{tabular}{lll}
$Ind\left( H\right) $ & $=$ & $+\frac{1}{2\pi }\left( \arctan \frac{B_{+}}{%
A_{+}}-\arctan \frac{B_{-}}{A_{+}}\right) +\frac{1}{2\pi }\left( \func{arccot%
}\frac{A_{+}}{B_{+}}-\func{arccot}\frac{A_{-}}{B_{+}}\right) $ \\
&  & $-\frac{1}{2\pi }\left( \arctan \frac{B_{+}}{A_{-}}-\arctan \frac{B_{-}%
}{A_{-}}\right) -\frac{1}{2\pi }\left( \func{arccot}\frac{A_{+}}{B_{-}}-%
\func{arccot}\frac{A_{-}}{B_{-}}\right) $%
\end{tabular}%
\end{equation}%
that coincides precisely with the Fukui formula (\ref{ind}); and which we
rewrite in the following remarkable factorised form
\begin{equation}
Ind\left( H\right) =\frac{1}{2}\left[ sgn\left( A_{+}\right) -sgn\left(
A_{-}\right) \right] \times \frac{1}{2}\left[ sgn\left( B_{+}\right)
-sgn\left( B_{-}\right) \right]  \label{ixy}
\end{equation}%
from which we directly learn the values of $Ind\left( H\right) $. This is a
topological relation; it depends only on the signature of the values of the
Higgs components at spatial infinities; a non zero index requires the non
vanishing of each factor in (\ref{ixy}) showing that $\phi _{x}\left(
x,y\right) $ has to change the sign when we go from $x\rightarrow -\infty $
to $x\rightarrow +\infty $ and the same thing should hold for $\phi
_{y}\left( x,y\right) $ when going from $y\rightarrow -\infty $ to $%
y\rightarrow +\infty $. This feature ensures that the closed curve contains
inside the pole singularity $\left \vert \mathbf{\phi }\right \vert =0$
where lives \textrm{gapless} states.

\section{Three dimensional BBH model}

The topological index of the three dimensional BBH lattice model with full
open boundary condition can be determined by studying the ground state
properties of a $4+4$ component fermion $\mathbf{\psi }=\mathbf{\psi }\left(
x,y,z\right) $ near the Dirac points. Around each one of these points of the
3D model, the fermion $\mathbf{\psi }$ is coupled to an external space
dependent field doublet $\phi ^{a}=\phi ^{a}\left( x,y,z\right) $ --- an $%
O\left( 3\right) $ Higgs field triplet--- with interacting dynamics
described by the typical matrix Hamiltonian
\begin{equation}
H=\mathbf{\Lambda }^{x}k_{x}+\mathbf{\Lambda }^{y}k_{y}+\mathbf{\Lambda }%
^{z}k_{z}+\mathbf{\Omega }^{x}\phi _{x}+\mathbf{\Omega }^{y}\phi _{y}+%
\mathbf{\Omega }^{z}\phi _{z}
\end{equation}%
In this relation, the six hermitian matrices $\mathbf{\Lambda }^{i}$\ and $%
\mathbf{\Omega }^{a}$\ are 8$\times $8 Dirac matrices realised in terms of
three sets of Pauli matrices $\mathbf{\sigma },\mathbf{\tau },\mathbf{\rho }$
as follows%
\begin{equation}
\begin{tabular}{lllllll}
$\mathbf{\Lambda }_{1}$ & $=$ & $\rho _{0}\otimes \tau _{2}\otimes \sigma
_{1}$ & $\quad ,\quad $ & $\mathbf{\Omega }_{1}$ & $=$ & $\rho _{0}\otimes
\tau _{2}\otimes \sigma _{2}$ \\
$\mathbf{\Lambda }_{2}$ & $=$ & $\rho _{0}\otimes \tau _{2}\otimes \sigma
_{3}$ & $\quad ,\quad $ & $\mathbf{\Omega }_{2}$ & $=$ & $\rho _{0}\otimes
\tau _{1}\otimes \sigma _{0}$ \\
$\mathbf{\Lambda }_{3}$ & $=$ & $-\rho _{2}\otimes \tau _{3}\otimes \sigma
_{0}$ & $\quad ,\quad $ & $\mathbf{\Omega }_{3}$ & $=$ & $-\rho _{1}\otimes
\tau _{1}\otimes \sigma _{0}$%
\end{tabular}
\label{mat}
\end{equation}%
From these anticommuting matrices that generate a Clifford algebra in 6D,
one define as well a chiral operator $\Gamma _{7}$ by the product $\frac{1}{i%
}\mathbf{\Lambda }_{1}\mathbf{\Omega }_{1}\mathbf{\Lambda }_{2}\mathbf{%
\Omega }_{2}\mathbf{\Lambda }_{3}\mathbf{\Omega }_{3}$ which reads in terms
of the Pauli matrices as follows%
\begin{equation}
\Gamma _{7}=\rho _{3}\otimes \mathbf{\gamma }_{5}=-\rho _{3}\otimes \tau
_{3}\otimes \sigma _{0}
\end{equation}%
This operator characterises the corner states which are described by chiral
wave functions. From the eigenvalue $H\mathbf{\psi }=E\mathbf{\psi },$ one
learns that $\mathbf{\psi }$ has $4+4$ components that can be formulated in
various ways; for example like the tensor product of three two component
spinors $\xi \otimes \zeta \otimes \eta $ as done in \cite{4F}; or simply
like $\mathbf{\psi }=\left( \mathbf{\lambda ,\bar{\chi}}\right) ^{T}$ with
components%
\begin{equation}
\mathbf{\lambda }=\left(
\begin{array}{c}
\lambda _{1} \\
\lambda _{2} \\
\lambda _{3} \\
\lambda _{4}%
\end{array}%
\right) \qquad ,\qquad \mathbf{\chi }=\left(
\begin{array}{c}
\chi _{1} \\
\chi _{2} \\
\chi _{3} \\
\chi _{4}%
\end{array}%
\right)
\end{equation}%
The corner states are determined by solving two constraint relations
generalising the previous ones in 2D; these are the gapless eigenvalue
equation $H\mathbf{\psi }=0$ with
\begin{equation}
H=-i\sum_{j=1}^{3}\mathbf{\Lambda }^{j}\frac{\partial }{\partial x^{j}}+%
\mathbf{\Omega }^{a}\phi _{a}  \label{30}
\end{equation}%
and the chirality condition
\begin{equation}
\Gamma _{7}\mathbf{\psi }=q_{_{\mathbf{\Gamma }}}\mathbf{\psi }  \label{33}
\end{equation}%
with chiral charge $q_{_{\mathbf{\Gamma }}}$ taking either the value $+1$ or
$-1$ and which is fixed by the normality condition of the corner state.

\subsection{Integral formula for Ind(H) in 3D}

The Hamiltonian index of the 3D model (\ref{30}) is given by the number of
chiral zero modes $N_{+}-N_{-}$ in the ground state. To calculate it, we
proceed in a quite similar manner as we have done in 2D model. First, we
think about this integer number as the flux of a three dimensional current $%
J^{i}=J^{i}\left( \mathbf{r}\right) $ like%
\begin{equation}
Ind\left( H\right) =\dint \nolimits_{\mathcal{S}}\vec{J}.\overrightarrow{dS}%
=\dint \nolimits_{\mathcal{S}}J^{i}dS_{i}  \label{js}
\end{equation}%
where $\overrightarrow{dS}$ is a vector surface element and $\mathcal{S}$ a
closed surface in the x-y-z space $\mathbb{R}_{\mathbf{r}}^{3}$ traversed by
the flux of $\vec{J}$. We can equivalently express the above relation as
follows%
\begin{equation}
Ind\left( H\right) =\dint \nolimits_{\mathcal{S}}\varpi  \label{om}
\end{equation}%
where now $\varpi $ is 2-form in the real 3D space%
\begin{equation}
\varpi =\frac{1}{2}\varpi _{jl}dx^{j}\wedge dx^{l}
\end{equation}%
related to the vector current $J^{i}$ like $\varpi _{jl}=\varepsilon
_{jli}J^{i}$ with $\varepsilon _{jli}$ the completely antisymmetric
Levi-Civita tensor in 3D. By substituting, we obtain%
\begin{equation}
Ind\left( H\right) =\frac{1}{2}\dint \nolimits_{\mathcal{S}}\varepsilon
_{ijl}J^{i}dx^{j}\wedge dx^{l}  \label{m}
\end{equation}%
Having introduced the integral formula for $Ind\left( H\right) $, we come
now to describe the vector current $J^{i}=\left \langle
J_{7}^{i}\right
\rangle $. It is given by the vacuum expectation value of
an axial vector current $J_{7}^{i}$ involving the $\mathbf{\Gamma }_{7}$
matrix as shown on the following expression%
\begin{equation}
J^{i}\left( \mathbf{r}\right) =\lim_{\mathbf{r}^{\prime }\mathbf{\rightarrow
r}}\left[ \lim_{m\rightarrow 0}tr\left( \mathbf{\Gamma }_{7}\mathbf{\Lambda }%
^{i}\frac{1}{m+iH}\right) \mathbf{\delta }_{3}\left( \mathbf{r}-\mathbf{r}%
^{\prime }\right) \right]  \label{jj}
\end{equation}%
that descends from an underlying quantum field theory description where $%
J_{7}^{i}=\mathbf{\psi }^{\dagger }\mathbf{\mathbf{\Gamma }_{7}\mathbf{%
\Lambda }^{i}\psi }$. \textrm{The expression of the above current }$%
J^{i}\left( \mathbf{r}\right) $\textrm{\ can be also motivated from the two
dimensional analysis of section 2; in particular from the study done
subsection 2.1. The }$J^{i}\left( \mathbf{r}\right) $\textrm{\ in Eq(\ref{jj}%
) is nothing but the 3D- extension of the two dimensional current given by
Eq(\ref{j}).\ By comparing the two expressions (\ref{jj}) and (\ref{j}), one
learns amongst others the two following indicators: First, the 2D- Dirac-
delta function }$\mathbf{\delta }_{2}\left( \mathbf{r}-\mathbf{r}^{\prime
}\right) $\textrm{\ in (\ref{j}) ---with }$\mathbf{r}=\left( x,y\right) $---%
\textrm{\ has been promoted to the three dimensional }$\mathbf{\delta }%
_{3}\left( \mathbf{r}-\mathbf{r}^{\prime }\right) $\textrm{\ in (\ref{jj}%
)---with }$\mathbf{r}=\left( x,y,z\right) $---\textrm{. Second, the chiral
operator }$\mathbf{\gamma }_{5}$\textrm{\ in Eq(\ref{j}) has been also
promoted to the chiral operator }$\mathbf{\Gamma }_{7}$\  \textrm{of 3D
space. Recall that the usual vector space dimension d}$_{v}$\textrm{\ and
the spinor dimension d}$_{s}$\textrm{\ are related to each other like d}$%
_{s}=2^{d_{v}}$\textrm{. So, we have in 2D, we have d}$_{s}=2^{2}=4$ and
then $\mathbf{\gamma }_{5}$\textrm{\ is a 4}$\times $\textrm{4 matrix
realised in our study as }$-\tau _{3}\otimes \sigma _{0}$\textrm{. In three
dimensions, the spinor dimension d}$_{s}$\textrm{\ is equal to} $\mathrm{2}%
^{3}=8$\textrm{; then }$\mathbf{\Gamma }_{7}$\textrm{\ is a }$\mathrm{8}%
\times \mathrm{8}$ \textrm{matrix realised here as }$\rho _{3}\otimes
\mathbf{\gamma }_{5}$\textrm{.} By calculating the trace over the $8\times 8$
matrix product in (\ref{jj}) and taking the vanishing mass limit, the above
vector current can be expressed in terms of the Higgs field triplet as%
\begin{equation}
J^{i}=\frac{\varepsilon ^{abc}}{8\pi \left \vert \mathbf{\phi }\right \vert
^{3}}\varepsilon ^{ijl}\phi _{a}\partial _{j}\phi _{b}\partial _{l}\phi _{c}
\label{3j}
\end{equation}%
with $\left \vert \mathbf{\phi }\right \vert ^{2}=\phi _{x}^{2}+\phi
_{y}^{2}+\phi _{z}^{2}$. \textrm{An explicit derivation of} \textrm{Eq.(\ref%
{3j}) is as done in the appendix for the 2D case by following the same steps
described there and by using properties of the }$\mathbf{\Lambda }^{i}$\
\textrm{and} $\mathbf{\Omega }^{a}$\textrm{\ matrices (\ref{mat}) which are
induced by the usual features of Pauli matrices. }Putting this 3D vector
current $J^{i}$ back into the relation (\ref{m}), we get the following index
formula%
\begin{equation}
Ind\left( H\right) =\dint \nolimits_{\mathcal{S}}\frac{1}{8\pi \left \vert
\mathbf{\phi }\right \vert ^{2}}dS_{i}\varepsilon ^{ijl}\phi _{a}\partial
_{j}\phi _{b}\partial _{l}\phi _{c}\varepsilon ^{abc}  \label{mm}
\end{equation}%
which can be interpreted as the winding number of the three component Higgs
vector $\phi _{a}$ field around the closed surface $\mathcal{S}$.

\subsection{Topological index from 3D Higgs space}

To determine the Ind(H) value of the Hamiltonian (\ref{30}), we have to
perform the integral in Eq.(\ref{mm}). To that purpose, we will proceed as
follows: As a front matter, we give some useful properties of the relation (%
\ref{mm}) including two examples of shapes of the surface $\mathcal{S}$
appearing in the integral formula of Ind(H); these $\mathcal{S}$ shapes are
given by a 2-sphere $\mathbb{S}_{\mathbf{r}}^{2}$ at infinity; and a
parallelepiped surface $\mathcal{S}_{\infty }^{\left( 8\right) }$
respectively as in Eqs.(\ref{s2}-\ref{si}). Then, we use results from
differential geometry on $\mathbb{R}^{3}$ to map Eq(\ref{mm}) into an
equivalent formula for Ind(H) which is completely expressed in the Higgs
space. This new formula is given by%
\begin{equation}
Ind\left( H\right) =\dint \nolimits_{\Sigma }\frac{1}{8\pi \left \vert
\mathbf{\phi }\right \vert ^{3}}\phi _{a}d\phi _{b}\wedge d\phi
_{c}\varepsilon ^{abc}  \label{fm}
\end{equation}%
it involves a closed surface $\Sigma $ to be introduced later on and has a
singularity at $\left \vert \mathbf{\phi }\right \vert =0$. The above
formula, to be derived in what follows, plays an important role in our
forthcoming calculations.

\subsubsection{Irregular boundary surface\emph{\ }$\mathcal{S}$}

Roughly speaking, the $Ind\left( H\right) $ defined by Eq.(\ref{mm}) is a
function of $\mathcal{S}$; so the integral formula (\ref{mm}) can be defined
as
\begin{equation}
Ind\left( H\right) =\mathcal{F}\left[ \mathcal{S}\right]
\end{equation}%
This expression shows that if we want to find the numerical value of Ind(H),
we should specify $\mathcal{S}$; unless if the integral (\ref{mm}) is
independent of the shape of $\mathcal{S}$. This is precisely what happens in
the present case since the Ind(H) is a topological invariant meaning that $%
\mathcal{F}\left[ \mathcal{S}\right] =\mathcal{F}\left[ \mathcal{S}^{\prime }%
\right] $ for any $\mathcal{S}^{\prime }$ related to $\mathcal{S}$ by a
continuous transformation. In this topological change the $\mathcal{S}$ is
deformed to $\mathcal{S}^{\prime }$ without affecting the value of Ind(H).
Nevertheless, let us think of the closed surface in (\ref{mm}) as given by a
big $\mathcal{S}_{\infty }$ at the boundary of the x-y-z space $\mathbb{R}_{%
\mathbf{r}}^{2}$ as in the following Figure.
\begin{figure}[tbph]
\begin{center}
\hspace{0cm} \includegraphics[width=10cm]{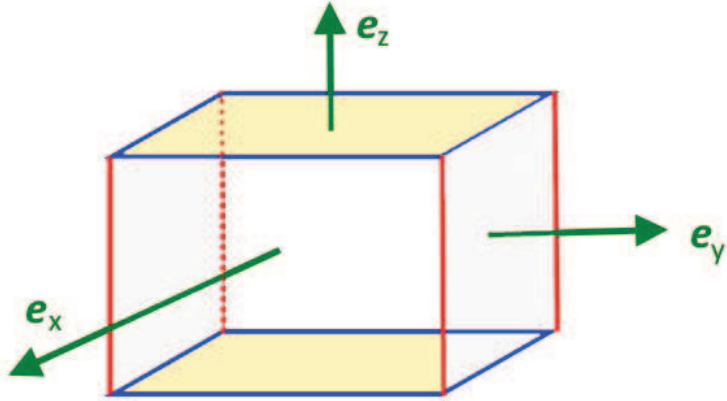}
\end{center}
\par
\vspace{-1 cm}
\caption{Boundary surface of a parallelepiped in 3D space given by 3+3
faces. These faces bi-intersect on twelve edges and tri-intersect at eight
tops. It corresponds to $\mathcal{S}^{\left( 8\right) }$ from the view of
the position space $\mathbb{R}_{\mathbf{r}}^{3}$ and to the surface\ $\Sigma
^{\left( 8\right) }$ from the view of $\mathbb{R}_{\mathbf{\protect \phi }%
}^{3}$.}
\label{c3}
\end{figure}
For concreteness, we consider below two $\mathcal{S}_{\infty }$ and $%
\mathcal{S}_{\infty }^{\prime }$ shapes as follows: $\left( i\right) $ $%
\mathcal{S}_{\infty }$ given by a 2-sphere $\mathbb{S}_{\mathbf{r}}^{2}$
with defining equation
\begin{equation}
\mathbb{S}_{\mathbf{r}_{\infty }}^{2}:x_{\infty }^{2}+y_{\infty
}^{2}+z_{\infty }^{2}=r_{\infty }^{2}  \label{s2}
\end{equation}%
it is the boundary of the ball $\mathcal{V}_{\infty }$ with equation $%
\left
\vert \mathbf{r}\right \vert <r_{\infty }$ ; so we have $\mathbb{S}_{%
\mathbf{r}}^{2}=\partial \mathcal{V}$; and $\left( ii\right) $ $\mathcal{S}%
_{\infty }^{\prime }$ given by the boundary surface $\mathcal{S}_{\infty
}^{\left( 8\right) }$ of cube --- or in general a regular parallelepiped---.
This cubic shape is an interesting situation that concerns 3D matter with
full open boundary conditions. In this case, $\mathcal{S}_{\infty }^{\left(
8\right) }$ delimits a volume $\mathcal{V}_{\infty }^{\left( 8\right) }$ and
is given by%
\begin{equation}
\mathcal{S}_{\infty }^{\left( 8\right) }=\left( \mathcal{S}_{1}^{+}\cup
\mathcal{S}_{1}^{-}\right) \cup \left( \mathcal{S}_{2}^{+}\cup \mathcal{S}%
_{2}^{-}\right) \cup \left( \mathcal{S}_{3}^{+}\cup \mathcal{S}%
_{3}^{-}\right)  \label{si}
\end{equation}%
Concretely $\mathcal{S}_{\infty }^{\left( 8\right) }$ has $3+3$ planar faces
$\mathcal{S}_{i}^{\pm }$; and eight tops with coordinates $\left(
x,y,z\right) =\left( \pm L_{x},\pm L_{y},\pm L_{z}\right) $. The planar $%
\mathcal{S}_{i}^{\pm }$ are normal to the x- y- z directions as depicted in
the \textbf{Figure (\ref{c3}); }they\ bi-intersect along the 12 following
segments
\begin{eqnarray}
\boldsymbol{C}_{x}^{\left( p,q\right) } &\sim &\mathcal{S}_{y}^{p}\cap
\mathcal{S}_{z}^{q}  \notag \\
\boldsymbol{C}_{y}^{\left( p,q\right) } &\sim &\mathcal{S}_{x}^{p}\cap
\mathcal{S}_{z}^{q}\qquad ,\qquad p,q=\pm \\
\boldsymbol{C}_{z}^{\left( p,q\right) } &\sim &\mathcal{S}_{x}^{p}\cap
\mathcal{S}_{y}^{q}  \notag
\end{eqnarray}%
and tri-intersect at the eight tops%
\begin{equation}
\boldsymbol{T}^{\left( p,q,s\right) }\sim \mathcal{S}_{1}^{p}\cap \mathcal{S}%
_{2}^{q}\cap \mathcal{S}_{3}^{s}\qquad ,\qquad p,q,s=\pm
\end{equation}%
This $\mathcal{S}_{\infty }^{\left( 8\right) }$ can be also interpreted as
describing the boundary of a cube ---parallelepiped--- circumscribed into a
2-sphere ---ellipsoid--- as depicted in the \textbf{Figure \ref{C4}}. With
this picture, one clearly see that one can pass from $\mathcal{S}_{\infty
}^{\left( 8\right) }$ to the 2-sphere $\mathbb{S}_{\infty }^{2}$ by
deforming the planar faces $\mathcal{S}_{i}^{\pm }$ into spherical calottes
with a square (rectangular) section. The surface shape (\ref{si}) will be
used later on when studying the 3D extension of the construction done for
the 2D model studied in the previous section.

\subsubsection{Deriving the signature Eq.(\protect \ref{A})}

\begin{figure}[tbph]
\begin{center}
\hspace{0cm} \includegraphics[width=8cm]{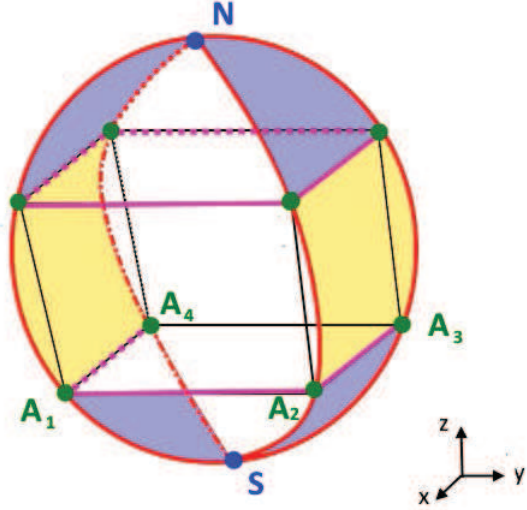}
\end{center}
\par
\vspace{-1 cm}
\caption{A boundary surface space of a cube circumscribed inside a 2-sphere $%
\mathbb{S}^{2}$. This picture holds both for the $\mathcal{S}_{\infty
}^{\left( 8\right) }$ of the position space and its image$\  \Sigma _{\infty
}^{\left( 8\right) }$ living in the Higgs space.}
\label{C4}
\end{figure}

A nice way to determine the Ind(H) given by Eq.(\ref{mm}) is to formulate it
directly in the Higgs space $\mathbb{R}_{\mathbf{\phi }}^{3}$ parameterised
by ($\phi _{x},\phi _{y},\phi _{z})$ and given by Eq.(\ref{fm}). For that,
we substitute the two following relations,
\begin{equation}
\begin{tabular}{lll}
$dS_{i}\varepsilon ^{ijl}$ & $=$ & $dx^{j}\wedge dx^{l}$ \\
$\  \ d\phi _{a}$ & $=$ & $\partial _{j}\phi _{a}dx^{j}$%
\end{tabular}%
\end{equation}%
back into Eq.(\ref{mm}) to end up exactly with (\ref{fm}). But, in this
index relation, the $\Sigma $ is a closed surface in the Higgs space ---$%
\Sigma $ is contained in $\mathbb{R}_{\mathbf{\phi }}^{3}$---; it is given
by the correspondence $\Phi :\mathcal{S}\rightarrow \Sigma $ mapping the
real surface $\mathcal{S}$ of the x-y-z space $\mathbb{R}_{\mathbf{r}}^{3}$
into the surface $\Sigma $ belonging $\phi _{x}$-$\phi _{y}$-$\phi _{z}$
space. Under this correspondence, each point $\mathbf{r}=\left( x,y,z\right)
$ in the 3D position space $\mathbb{R}_{\mathbf{r}}^{3}$ gets mapped into a
point $\mathbf{\phi }=\left( \phi _{x},\phi _{y},\phi _{z}\right) $ in the
3D Higgs space $\mathbb{R}_{\mathbf{\phi }}^{3}$. Notice that because of
this mapping, several relations in $\mathbb{R}_{\mathbf{r}}^{3}$ can be also
mapped into corresponding ones in $\mathbb{R}_{\mathbf{\phi }}^{3}$. Below,
we give two interesting examples respectively dealing with Eq.(\ref{js}) and
Eqs.(\ref{s2}-\ref{si}). The first example concerns the flux formula (\ref%
{js}) whose homologue in the Higgs space is obtained by setting%
\begin{equation}
d\sigma ^{a}=\frac{1}{2}\varepsilon ^{abc}d\phi _{b}\wedge d\phi _{c}
\end{equation}%
into Eq.(\ref{fm}); this substitution allows to bring the flux formula into
the interesting form%
\begin{equation}
ind\left( H\right) =\dint \nolimits_{\Sigma }\mathcal{B}_{a}d\sigma
^{a}=\dint \nolimits_{\Sigma }\mathcal{\vec{B}}.\overrightarrow{d\sigma }
\label{nf}
\end{equation}%
with vector field $\mathcal{B}_{a}=\frac{\phi _{a}}{4\pi \left \vert \mathbf{%
\phi }\right \vert ^{3}}$. This index relation, which describes the flux of $%
\mathcal{B}_{a}$ through $\Sigma $, is remarkable in the sense it can be
also put in correspondence with the well known Gauss theorem concerning the
electrostatic field of Coulomb theory. The second example concerns the
surface $\mathcal{S}$ and its image $\Sigma $; for $\mathcal{S}_{\infty }$
given by the 2-sphere $\mathbb{S}_{\mathbf{r}}^{2}$ of Eq.(\ref{s2}), it is
associated a 2-sphere $\Sigma _{\infty }=\mathbb{S}_{\mathbf{\phi }}^{2}$
given by
\begin{equation}
\mathbb{S}_{\mathbf{\phi }}^{2}:\phi _{x\infty }^{2}+\phi _{y\infty
}^{2}+\phi _{z\infty }^{2}=\phi _{\infty }^{2}
\end{equation}%
Similarly, for the $\mathcal{S}_{\infty }^{\left( 8\right) }$ given by (\ref%
{si}), we have the corresponding Higgs space surface
\begin{equation}
\Sigma _{\infty }^{\left( 8\right) }=\left( \Sigma _{1}^{+}\cup \Sigma
_{1}^{-}\right) \cup \left( \Sigma _{2}^{+}\cup \Sigma _{2}^{-}\right) \cup
\left( \Sigma _{3}^{+}\cup \Sigma _{3}^{-}\right)  \label{sg}
\end{equation}%
it has $3+3$ faces $\Sigma _{i}^{\pm }$ normal to the $\phi _{i}$-
directions in the Higgs space; twelve edges%
\begin{equation}
\mathcal{C}_{j}^{\left( p,q\right) }\sim \varepsilon _{lij}\Sigma
_{i}^{p}\cap \Sigma _{j}^{q}
\end{equation}%
and eight tops%
\begin{equation}
\mathcal{T}^{\left( p,q,s\right) }\sim \Sigma _{1}^{p}\cap \Sigma
_{2}^{q}\cap \Sigma _{3}^{s}\qquad ,\qquad p,q,s=\pm
\end{equation}%
with coordinates $\left( \phi _{x},\phi _{y},\phi _{z}\right) =\left( A_{\pm
},B_{\pm },C_{\pm }\right) $. Here also $\Sigma _{\infty }^{\left( 8\right)
} $ has an interpretation in terms of a cube ---parallelepiped---
circumscribed into the 2-sphere $\mathbb{S}_{\mathbf{\phi }}^{2}$
---ellipsoid--- as depicted in the \textbf{Figure \ref{C4}}. In this
realisation, the two parallel faces $\Sigma _{x}^{\pm }$ are normal to $\phi
_{x}$-direction and are located at $\phi _{x\rightarrow \pm \infty }=\phi
_{x_{\pm \infty }}$; a similar thing is valid for $\Sigma _{2}^{\pm }$ and $%
\Sigma _{3}^{\pm }$\ respectively normal to $\phi _{y}$- and $\phi _{z}$-
directions.
For later use notice that the planar faces $\Sigma _{i}^{\pm }$ can be also
defined as cross products of edges. Denoting the 12 edges of the cube
---parallelepiped---like%
\begin{eqnarray}
\mathcal{C}_{x}^{\left( p,q\right) } &:&A_{-}\leq \phi _{x}\leq A_{+}\quad
,\quad \left( \phi _{y},\phi _{z}\right) =\left( B_{p},C_{q}\right)  \notag
\\
\mathcal{C}_{y}^{\left( p,q\right) } &:&B_{-}\leq \phi _{y}\leq B_{+}\quad
,\quad \left( \phi _{x},\phi _{z}\right) =\left( A_{p},C_{q}\right) \\
\mathcal{C}_{z}^{\left( p,q\right) } &:&C_{-}\leq \phi _{z}\leq C_{+}\quad
,\quad \left( \phi _{x},\phi _{y}\right) =\left( A_{p},B_{q}\right)  \notag
\end{eqnarray}%
and the union of the four upper (resp. lower) edges of the face $\Sigma
_{3}^{+}$ (resp. $\Sigma _{3}^{-}$) as%
\begin{equation}
\mathcal{P}_{xy}^{\left( q\right) }=\mathcal{C}_{x}^{\left( +,q\right) }\cup
\mathcal{C}_{y}^{\left( +,q\right) }\cup \mathcal{C}_{x}^{\left( -,q\right)
}\cup \mathcal{C}_{y}^{\left( -,q\right) }
\end{equation}%
we can think of $\Sigma _{\infty }^{\left( 8\right) }$ as the union $\Sigma
_{L}\cup \left( \Sigma _{3}^{+}\cup \Sigma _{3}^{-}\right) $ with $\Sigma
_{L}$\ standing for the lateral surface $\left( \Sigma _{1}^{+}\cup \Sigma
_{1}^{-}\right) \cup \left( \Sigma _{2}^{+}\cup \Sigma _{2}^{-}\right) $ of
a cylinder with rectangular cross section. Moreover, using the relation $%
\Sigma _{L}\sim \mathcal{C}_{z}\times \mathcal{P}_{xy}$, we end up with%
\begin{equation}
\Sigma _{\infty }^{\left( 8\right) }\sim \left( \mathcal{C}_{z}\times
\mathcal{P}_{xy}\right) \cup \left( \Sigma _{3}^{+}\cup \Sigma
_{3}^{-}\right)  \label{s8}
\end{equation}%
With these data on the surface $\mathcal{S}_{\infty }$ and its image $\Sigma
_{\infty }$ in the Higgs space, we are now in position to determine the
value of Ind(H) by using the Higgs space formula (\ref{nf}). For the
spherical choice $\Sigma _{\infty }=\mathbb{S}_{\phi }^{2}$, it is
interesting to use following change of field variable,%
\begin{equation}
\begin{tabular}{lll}
$\phi _{x}$ & $=$ & $\phi \sin \vartheta \cos \mathrm{\varphi }$ \\
$\phi _{y}$ & $=$ & $\phi \sin \vartheta \sin \mathrm{\varphi }$ \\
$\phi _{z}$ & $=$ & $\phi \cos \vartheta $%
\end{tabular}
\label{hc}
\end{equation}%
to perform the integral in (\ref{nf}). From this change, we learn the
associated quantities $\varrho ^{2}=\phi _{x}^{2}+\phi _{y}^{2}$ and $\phi
^{2}=\varrho ^{2}+\phi _{z}^{2}$ as well as
\begin{equation}
\tan \varphi =\frac{\phi _{y}}{\phi _{x}}\quad ,\quad \cos \vartheta =\frac{%
\phi _{z}}{\phi }\quad ,\quad \tan ^{2}\vartheta =\frac{\varrho ^{2}}{\phi
^{2}}
\end{equation}%
Putting the field change (\ref{hc}) back into Eq.(\ref{nf}), we obtain%
\begin{equation}
Ind\left( H\right) =\pm \frac{1}{4\pi }\dint \nolimits_{\Sigma _{\infty
}}\sin \vartheta d\vartheta d\varphi
\end{equation}%
giving $\pm 1$; thanks to the radial symmetry of $\mathcal{B}_{a}$. For the
cartesian shape $\Sigma _{\infty }^{\left( 8\right) }$ given by (\ref{sg}),
it is interesting to still use the $\left( \vartheta ,\varphi \right) $
angles to compute the index $\int_{\Sigma _{\infty }^{\left( 8\right) }}%
\mathcal{B}_{a}d\sigma ^{a}$ which, as shown on (\ref{s8}), decomposes as
the sum of three terms like $\mathcal{J}_{0}+\mathcal{J}_{+}+\mathcal{J}_{-}$
with $\mathcal{J}_{0,\pm }$ respectively given by%
\begin{equation}
\mathcal{J}_{0}=\dint \nolimits_{\mathcal{C}_{z}\times \mathcal{P}_{xy}}%
\mathcal{B}_{a}d\sigma ^{a}\qquad ,\qquad \mathcal{J}_{\pm }=\dint
\nolimits_{\Sigma _{3}^{\pm }}\mathcal{B}_{a}d\sigma ^{a}
\end{equation}%
Straightforward calculations lead to
\begin{eqnarray}
\mathcal{J}_{0} &=&\frac{\mathcal{I}_{\mathcal{P}_{xy}}}{2}\left( \cos
\vartheta _{+}-\cos \vartheta _{-}\right)  \notag \\
\mathcal{J}_{+} &=&\frac{\mathcal{I}_{\mathcal{P}_{xy}}}{2}\left( 1-\cos
\vartheta _{+}\right)  \label{j3} \\
\mathcal{J}_{-} &=&\frac{\mathcal{I}_{\mathcal{P}_{xy}}}{2}\left( \cos
\vartheta _{-}+1\right)  \notag
\end{eqnarray}%
with $\vartheta _{\pm }=\arccos \frac{C_{\pm }}{\phi }$ and $\mathcal{I}_{%
\mathcal{P}_{xy}}$ as follows%
\begin{equation}
\mathcal{I}_{\mathcal{P}_{xy}}=\frac{1}{4}\left[ sgn\left( A_{+}B_{+}\right)
+sgn\left( A_{-}B_{-}\right) \right] -\frac{1}{4}\left[ sgn\left(
A_{+}B_{-}\right) +sgn\left( A_{-}B_{+}\right) \right]  \label{pxy}
\end{equation}%
The sum of the three relations of Eq.(\ref{j3}) gives exactly $\mathcal{I}_{%
\mathcal{P}_{xy}}$; it looks as it doesn't depend on sgn$\left( C_{\pm
}\right) $; and, according to (\ref{ixy}), it is an integer. A way to
interpret the absence of sgn$\left( C_{\pm }\right) $ is because of the
choice we have used in our calculation namely $sgn\left( C_{-}\right)
=-sgn\left( C_{+}\right) =-1.$ To implement, the contribution of sgn$\left(
C_{\pm }\right) $, we think of the above index as given by%
\begin{equation}
Ind(H)=\frac{\mathcal{I}_{\mathcal{P}_{xy}}}{2}\left[ sgn\left( C_{+}\right)
-sgn\left( C_{-}\right) \right]
\end{equation}%
by setting $sgn\left( C_{+}\right) =1,$ one recovers (\ref{pxy}). This
relation can be also motivated by its factorised form (\ref{ixy}) and cyclic
symmetry properties allowing to determine $\mathcal{I}_{\mathcal{P}_{yz}}$
and $\mathcal{I}_{\mathcal{P}_{zx}}$ from $\mathcal{I}_{\mathcal{P}_{xy}}$.
Expanding the above generalisation, we get%
\begin{equation}
\begin{tabular}{lll}
$Ind(H)$ & $=$ & $+\frac{1}{8}\left[ sgn\left( A_{+}B_{+}C_{+}\right)
+sgn\left( A_{-}B_{-}C_{+}\right) \right] $ \\
&  & $+\frac{1}{8}\left[ sgn\left( A_{+}B_{-}C_{-}\right) +sgn\left(
A_{-}B_{+}C_{-}\right) \right] $ \\
&  & $-\frac{1}{8}\left[ sgn\left( A_{+}B_{-}C_{+}\right) +sgn\left(
A_{-}B_{+}C_{+}\right) \right] $ \\
&  & $-\frac{1}{8}\left[ sgn\left( A_{+}B_{+}C_{-}\right) +sgn\left(
A_{-}B_{-}C_{-}\right) \right] $%
\end{tabular}
\label{ih3}
\end{equation}%
having eight contributions in one to one correspondence with the corners of
the cube ---parallelepiped--- with tops $\left( A_{\pm },B_{\pm },C_{\pm
}\right) $. In the end notice that, like for (\ref{ixy}) of the 2D model,
this topological formula factorises as follows,
\begin{eqnarray}
Ind(H) &=&\frac{1}{2}\left[ sgn\left( A_{+}\right) -sgn\left( A_{-}\right) %
\right] \times  \notag \\
&&\frac{1}{2}\left[ sgn\left( B_{+}\right) -sgn\left( B_{-}\right) \right]
\times  \label{3ih} \\
&&\frac{1}{2}\left[ sgn\left( C_{+}\right) -sgn\left( C_{-}\right) \right]
\notag
\end{eqnarray}%
and showing that here also that $sgn\left( \phi _{a}\right) $ has to change
its polarity when we go from minus infinity to plus infinity in order to
have a non trivial value of the index. For a non trivial value of Ind(H),
the surface $\Sigma $\ has to englobe the pole singularity at $\left \vert
\mathbf{\phi }\right \vert =0$.

\section{Conclusion and comments}

In this paper, we studied the topological properties of the three
dimensional BBH lattice model falling into the DBI class in the AZ periodic
table with reflection symmetries. First, we revisited the 2D model and
re-derived the topological index Ind(H$_{2D})$ of this theory by using
topological mapping from the real x-y plane $\mathbb{R}_{\mathbf{r}}^{2}$
into the $\phi _{x}$-$\phi _{y}$ Higgs plane $\mathbb{R}_{\mathbf{\phi }%
}^{2} $. Then, we investigated the topological Ind(H$_{3D})$ of the 3D
theory and calculated its expression in terms of the limit values of the
Higgs field triplet at space infinity. The topological index formula given
by (\ref{ih3}) can remarkably factorise like in (\ref{3ih}). Our method
revealed that the results obtained for 2D, given by (\ref{ind};\ref{ixy});
and their homologue derived for 3D, correspond to leading terms of a general
DBI topological index formula given by%
\begin{equation}
Ind(H_{ND})=\dprod \limits_{a=1}^{N}\frac{1}{2}\left[ sgn\left(
A_{+a}\right) -sgn\left( A_{-a}\right) \right]
\end{equation}%
where $\left( A_{\pm i}\right) _{1\leq i\leq N}$'s stand for the values at
space infinities of a N- component Higgs field multiplet; i.e $\mathbf{\phi }%
=\left( \phi _{a}\right) _{1\leq a\leq N}$. Non zero topological index
requires $A_{+a}$ and $A_{-a}$ to have opposite signs. The above relation
depends only on the signatures of the $A_{p_{a}}$ values with $p_{a}=\pm 1$;
and can be also expressed in other different ways: for example like%
\begin{equation}
Ind(H_{ND})=\dprod \limits_{a=1}^{N}\sum_{p_{a}=\pm }\frac{p_{a}}{2}%
sgn\left( A_{p_{a}}\right)
\end{equation}%
By setting $sgn\left( A_{\pm _{a}}\right) =\left( -\right) ^{\xi _{\pm a}}$
with integer $\xi _{\pm a}=0,1$ $\func{mod}2$, the above index formula can
be re-expressed as follows%
\begin{equation}
Ind(H_{ND})=\dprod \limits_{a=1}^{N}\frac{1}{2}\left( e^{i\pi \xi
_{+a}}-e^{i\pi \xi _{-a}}\right)
\end{equation}%
which, up to a sign, reads like $\frac{1}{2^{N}}\dprod
\nolimits_{a=1}^{N}\left( 1-e^{i\pi \xi _{a}}\right) $ with $\xi _{a}$\
given by the difference $\xi _{+a}-\xi _{-a}$. Like for 2D and 3D, a non
zero value of the Hamiltonian index in higher dimensions requires that the
boundary hypersurface $\Sigma $ contains in its inside the pole singularity $%
\left \vert \mathbf{\phi }\right \vert =0$.

\section{Appendices}

Here, we give two appendices A and B aiming some technical details, which
for simplicity of the presentation and also for the chain of ideas, have
been omitted in the heart of the paper. In appendix A, we give some useful
information on\ the Hamiltonians (\ref{01}-\ref{02}) and make a comment
regarding the vector field $\left( \phi _{x},\phi _{y},\phi _{z}\right) $.
In appendix B, we give explicit details regarding the derivation of the 2D
topological current of Eq.(\ref{cu}) used in sections 3 and 4.

\subsection{Appendix A: More on Hamiltonian (\protect \ref{01})}

The 3D- extension of the two dimensional BBH lattice Hamiltonian model with
full open boundary condition reads in reciprocal space as follows
\begin{equation}
\begin{tabular}{lll}
$\boldsymbol{H}_{lat}$ & $=$ & $+t_{x}\left( \sin k_{x}\right) \mathbf{%
\Lambda }^{x}+\left( \Delta _{x}+t_{x}\cos k_{x}\right) \mathbf{\Omega }^{x}$
\\
&  & $+t_{y}\left( \sin k_{y}\right) \mathbf{\Lambda }^{y}+\left( \Delta
_{y}+t_{y}\cos k_{y}\right) \mathbf{\Omega }^{y}$ \\
&  & $+t_{z}\left( \sin k_{z}\right) \mathbf{\Lambda }^{z}+\left( \Delta
_{z}+t_{z}\cos k_{z}\right) \mathbf{\Omega }^{z}$%
\end{tabular}
\label{la}
\end{equation}%
where $\mathbf{\Lambda }^{i}$\ and $\mathbf{\Omega }^{i}$ are hermitian $%
8\times 8$ matrices given by Eqs(\ref{mat}) and where $\Delta _{i}$ and $%
t_{i}$ are hopping parameters: intra and extra unit cells. For simplicity of
the presentation given below, we set $t_{x}=t_{y}=t_{z}=1$; the reduction
down to 2D can be obtained by cutting the z-direction ($t_{z}=\Delta _{z}=0$%
) and thinking of the remaining reduced $\mathbf{\Lambda }^{i}$\ and $%
\mathbf{\Omega }^{i}$ as hermitian $4\times 4$ matrices with realisation as
in Eqs(\ref{ma}). Notice that by setting $\Theta _{i}=\left( \Delta
_{i}+\cos k_{i}\right) $ with $i=x,y,z$, we can express the above lattice
Hamiltonian $\boldsymbol{H}_{lat}$ as a matrix function of the 3 momentum
variables $\left( k_{x},k_{y},k_{z}\right) $ and the three $\left( \Theta
_{x},\Theta _{y},\Theta _{z}\right) $; that is
\begin{equation}
\boldsymbol{H}_{lat}=\boldsymbol{H}_{lat}\left( k_{x},k_{y},k_{z};\Theta
_{x},\Theta _{y},\Theta _{z}\right) .
\end{equation}%
However as $\Theta _{i}=\Theta \left( k_{i}\right) ,$ the lattice
Hamiltonian is then a function $\boldsymbol{H}_{lat}\left(
k_{x},k_{y},k_{z}\right) ;$ so the symmetry constraints (\ref{c}) and (\ref%
{d}) apply as well to $\boldsymbol{H}_{lat}=\sum_{i}(\mathbf{\Lambda }%
^{i}\sin k_{i}+\mathbf{\Omega }^{i}\Theta _{i})$. To see the relationship
between this $\boldsymbol{H}_{lat}$ and (\ref{01}); we calculate the eight
energy egenvalues of $\boldsymbol{H}_{lat};$ they are nicely obtained by
computing $\boldsymbol{H}_{lat}^{2}$ which turns out to be proportional to
the identity matrix $I_{8\times 8}$; that is $\boldsymbol{H}%
_{lat}^{2}=E_{lat}^{2}I_{8\times 8}$. This feature leads to the two
following $E_{lat}^{\pm }$ energies with multiplicity of order 4,%
\begin{equation}
E_{lat}^{\pm }=\pm \sqrt{\sum_{i=x,y,z}\left[ \sin ^{2}k_{i}+\left( \Delta
_{i}+\cos k_{i}\right) ^{2}\right] }.
\end{equation}%
The gap energy is determined by looking for the minimal value of $\left(
E_{lat}^{+}\right) _{\min }$; it is obtained by solving the vanishing of two
following sets of equations:
\begin{equation}
\left( I\right) :\sin k_{i}=0\qquad ,\qquad \left( II\right) :\Delta
_{i}+\cos k_{i}=0.
\end{equation}%
While the set $\left( II\right) $ shows that for $\left \vert \Delta
_{i}\right \vert >1$, the system is gapped; the first set teaches us
interesting information. First, it has eight solutions that can be expressed
like $k_{i}^{\ast }=n_{i}\pi $ with $n_{i}=0,1$ mod 2. Therefore, given a
fix point $\left( k_{x}^{\ast },k_{y}^{\ast },k_{z}^{\ast }\right) =\left(
n_{x}\pi ,n_{y}\pi ,n_{z}\pi \right) $ with some integers $\left(
n_{x},n_{y},n_{z}\right) ,$ the momentum vector $k_{i}$ around the $%
k_{i}^{\ast }$ expands as follows%
\begin{equation}
k_{i}\simeq k_{i}^{\ast }+k_{i}^{\prime }=n_{i}\pi +k_{i}^{\prime }
\end{equation}%
where $k_{i}^{\prime }$ is a small deviation; that is $\left \vert \mathbf{k}%
^{\prime }\right \vert /\left \vert \mathbf{k}^{\ast }\right \vert <<1$.
Second, putting this change back into in Eq(\ref{la}), we can approximate
the lattice Hamiltonian $\boldsymbol{H}_{lat}\left( \mathbf{k}\right) $ near
$\mathbf{k}^{\ast }$ like $H_{n_{x},n_{y},n_{z}}\left( \mathbf{k}^{\prime
}\right) +O\left( \mathbf{k}^{\prime 2}\right) $ with
\begin{equation}
H_{n_{x},n_{y},n_{z}}=\sum_{i=x,y,z}{\small (-)}^{n_{i}}\left[ k_{i}^{\prime
}\mathbf{\Lambda }^{i}+\varphi _{n_{i}}^{\prime }\mathbf{\Omega }^{i}\right]
\end{equation}%
with $\varphi _{n_{i}}^{\prime }=1+\Delta _{i}\cos \left( n_{i}\pi \right) $%
. For example, near the point $\left( k_{x}^{\ast },k_{y}^{\ast
},k_{z}^{\ast }\right) $ given by the origin $\left( 0,0,0\right) $, we have
the Hamiltonian $H_{0,0,0}=\sum_{i}\left[ k_{i}^{\prime }\mathbf{\Lambda }%
^{i}+\varphi _{i}\mathbf{\Omega }^{i}\right] $ with $\varphi _{i}^{\prime
}=\left( 1+\Delta _{i}\right) $. Quantum fluctuations are described by
realising the momentum deviations $k_{i}^{\prime }$ in terms of the
operators $k_{x}^{\prime }=-i\frac{\partial }{\partial x^{\prime }}$ and so
on. Topological properties are described by promoting the above $\varphi
_{i}^{\prime }$'s space coordinate dependent Higgs $\phi _{i}(\mathbf{r}%
^{\prime })$ with $\mathbf{r}^{\prime }=\left( x^{\prime },y^{\prime
},z^{\prime }\right) $. By dropping out the primes from $\mathbf{k}^{\prime
} $ and $\mathbf{r}^{\prime };$ one recovers amongst others Eqs. (\ref{01}-%
\ref{02}).

\subsection{Appendix B: Derivation of Eq.(\protect \ref{cu})}

Here, we give a rapid sketch of the derivation of the topological current $%
J^{i}\left( \mathbf{r}\right) $ given by Eq.(\ref{cu}) by starting from the
following definition of the axial current using Pauli-Villars regularisation
parameter M,%
\begin{equation}
J^{i}\left( \mathbf{r}\right) =\lim_{\mathbf{r}^{\prime }\mathbf{\rightarrow
r}}\lim_{m\rightarrow 0}\lim_{M\rightarrow \infty }tr\left[ \mathbf{\gamma }%
_{5}\mathbf{\Lambda }^{i}\left( \frac{1}{iH+m}-\frac{1}{iH+M}\right) \mathbf{%
\delta }_{2}\left( \mathbf{r}-\mathbf{r}^{\prime }\right) \right] .
\label{a1}
\end{equation}%
Strictly speaking, this relation involves four operations that we have to
perform in order to put $J^{i}\left( \mathbf{r}\right) $ into the remarkable
form (\ref{cu}) used in the paper. These operations are given by the matrix
trace $tr\left( ...\right) $ which we have to calculate; and moreover three
limits namely $\lim_{M\rightarrow \infty }$ and $\lim_{m\rightarrow 0}$ as
well as $\lim_{\mathbf{r}^{\prime }\mathbf{\rightarrow r}}$ that we have to
perform as well. In our present situation, the limit $\lim_{M\rightarrow
\infty }$ is trivial and the term 1/$\left( iH+M\right) $ in (\ref{a1}) can
be dropped out. So, we are left with the basic term that we express like%
\begin{equation}
J^{i}\left( \mathbf{r}\right) =\lim_{\mathbf{r}^{\prime }\mathbf{\rightarrow
r}}\lim_{m\rightarrow 0}tr\left[ \mathbf{\gamma }_{5}\mathbf{\Lambda }%
^{i}\left( \frac{m-iH}{H^{2}+m^{2}}\right) \mathbf{\delta }_{2}\left(
\mathbf{r}-\mathbf{r}^{\prime }\right) \right] .
\end{equation}
To proceed, we first perform the $\lim_{\mathbf{r}^{\prime }\rightarrow
\mathbf{r}}$ by substituting the Dirac- delta function $\mathbf{\delta }%
_{3}\left( \mathbf{r-r}^{\prime }\right) $ by its expression as an integral
over the plane waves $e^{i\mathbf{k.}\left( \mathbf{r-r}^{\prime }\right) }$
and think of $H$ like a differential operator $-i\Lambda ^{j}\partial
_{j}+\Omega ^{a}\phi _{a}$; this leads to put $J^{i}\left( \mathbf{r}\right)
$ in the form $\lim_{m\rightarrow 0}\mathcal{J}^{i}\left( \mathbf{r;}%
m\right) $ with
\begin{equation}
\mathcal{J}^{i}\left( \mathbf{r;}m\right) =\dint \nolimits_{R^{2}}\frac{d^{2}%
\mathbf{k}}{\left( 2\pi \right) ^{2}}tr\left[ \mathbf{\gamma }_{5}\mathbf{%
\Lambda }^{i}e^{-i\mathbf{k.r}}\left( \frac{m-iH}{H^{2}+m^{2}}\right) e^{i%
\mathbf{k.r}}\right]  \label{ki}
\end{equation}%
We will show later on that the $\lim_{m\rightarrow 0}$ is also trivial here;
but let us keep it for a moment and kill it at proper time. The remaining
steps to do are technical and rely on some key ingredients; in particular
the two following ones that we want to comment on as they concern crucial
stages: $\left( i\right) $ The way to deal with the inverse operator $\frac{1%
}{H^{2}+m^{2}}$ appearing in (\ref{ki}) as it hides a difficulty that we
have to overcome in order to get a simple expression of the current. The
point is that the quantity $e^{-i\mathbf{k.r}}H^{2}e^{+i\mathbf{k.r}}$,
which is equal to $\mathbf{k}^{2}+\mathbf{\phi }^{2}-i\Lambda ^{j}\Omega
^{a}\partial _{j}\phi _{a}+\mathcal{O}_{2}$ with $\mathcal{O}_{2}=-\mathbf{%
\nabla }^{2}-2i\left( \mathbf{k.\nabla }\right) $, involves the field matrix
$\Lambda ^{j}\Omega ^{a}\partial _{j}\phi _{a}$. After substituting, we can
put $e^{-i\mathbf{k.r}}\left( H^{2}+m^{2}\right) ^{-1}e^{+i\mathbf{k.r}}$ as
the product of $\left( \mathbf{k}^{2}+\mathbf{\phi }^{2}+m^{2}\right) ^{-1}$
with $1/\left( I\mathbf{+}\Theta _{m}\right) $ where $\Theta _{m}$ is
roughly given by the matrix $-i\Lambda ^{j}\Omega ^{a}\partial _{j}\phi _{a}$
divided by the number $\left( \mathbf{k}^{2}+\mathbf{\phi }^{2}+m^{2}\right)
$; that is%
\begin{equation}
\Theta _{m}=\frac{-i\Lambda ^{j}\Omega ^{a}\partial _{j}\phi _{a}+\mathcal{O}%
_{2}}{\mathbf{k}^{2}+\mathbf{\phi }^{2}+m^{2}}
\end{equation}%
In other words, the term $\left( H^{2}+m^{2}\right) ^{-1}$ in (\ref{ki})
involves the fraction $1/\left( I\mathbf{+}\Theta _{m}\right) $ having
matrices in the denominator. But this dependence poses a problem when coming
to the explicit calculation of the matrix trace in (\ref{ki}). Hopefully,
one can use perturbation theory methods to replace $1/\left( I\mathbf{+}%
\Theta _{m}\right) $ by its expansion in $\Theta $- powers given by $%
I-\Theta _{m}+$ higher powers. This demands however assuming $\left \vert
\Theta \right \vert $ small which is equivalent the condition $\left \vert
\partial _{j}\phi _{a}\right \vert <<\left \vert \mathbf{\phi }\right \vert
^{2} $ requiring a slow variation of the field gradient $\nabla \mathbf{\phi
}$ with respect to the variation of $\mathbf{\phi }$. $\left( ii\right) $
The second key ingredient we want to comment comes from the computation of
matrix traces like $tr\left[ \mathbf{\gamma }_{5}\mathbf{\Lambda }^{i}\left(
X\right) \right] $ where X is given by products type $\Pi _{n=1}^{N}\Lambda
^{i_{n}}\Pi _{n^{\prime }=1}^{N^{\prime }}\Omega ^{j_{n^{\prime }}}$ and
having in mind that $tr\left( \Lambda ^{i}\right) =tr\left( \Omega
^{i}\right) =0$ and $tr\left( I_{4}\right) =4$. However, traces of the form $%
tr\left[ \mathbf{\gamma }_{5}\mathbf{\Lambda }^{i}\left( X\right) \right] $
have non vanishing values except for $X\equiv X_{i}$ proportional to the
product $\varepsilon _{ij}\mathbf{\Lambda }^{j}\times \left( \mathbf{\Omega }%
^{k}\mathbf{\Omega }^{l}\varepsilon _{kl}\right) $. In this regards, recall
that $\mathbf{\gamma }_{5}=-\mathbf{\Lambda }^{x}\mathbf{\Omega }^{x}\mathbf{%
\Lambda }^{y}\mathbf{\Omega }^{y}$ and so $tr\left( \mathbf{\gamma }_{5}%
\mathbf{\Lambda }^{x}\mathbf{\Lambda }^{y}\mathbf{\Omega }^{x}\mathbf{\Omega
}^{y}\right) =4$. This means that the $\frac{m-iH}{H^{2}+m^{2}}$ in (\ref{ki}%
) should contribute by $X_{i}$ given by the product of three matrices: one
appropriate $\mathbf{\Lambda }$ and two $\mathbf{\Omega }$'s. This feature
rules out the term $\frac{m}{H^{2}+m^{2}}$ in (\ref{ki}) leaving only $\frac{%
-iH}{H^{2}+m^{2}}$. This is because the monomials $\left[ \Theta _{m}\right]
^{n}$ in the expansion of $1/\left( I\mathbf{+}\Theta _{m}\right) $ produces
terms like $\left( \Lambda ^{j}\Omega ^{a}\right) ^{n}$ while a non
vanishing trace requires terms as $\varepsilon _{ij}\mathbf{\Lambda }%
^{j}\times \left( \mathbf{\Omega }^{a}\mathbf{\Omega }^{b}\varepsilon
_{ab}\right) $. By taking the limit $\lim_{m\rightarrow 0}$ , we can put (%
\ref{ki}) into the following factorised form
\begin{equation}
\mathcal{J}^{i}\left( \mathbf{r}\right) =tr\left( \mathbf{\gamma }_{5}%
\mathbf{\Lambda }^{i}\Omega ^{a}\Lambda ^{j}\Omega ^{b}\right) \times \phi
_{a}\partial _{j}\phi _{b}\times \dint \nolimits_{R^{2}}\frac{d^{2}\mathbf{k}%
}{\left( 2\pi \right) ^{2}}\frac{1}{\left( \mathbf{k}^{2}+\mathbf{\phi }%
^{2}\right) ^{2}}  \label{jk}
\end{equation}%
where we have dropped out irrelevant terms such as those involving the $%
\mathcal{O}_{i}^{\prime }$s. By substituting the trace $tr\left( \mathbf{%
\gamma }_{5}\mathbf{\Lambda }^{i}\Omega ^{a}\Lambda ^{j}\Omega ^{b}\right) $
by its value $4\varepsilon ^{ij}\varepsilon ^{ab}$; then performing the
change $\mathbf{q}=\mathbf{k}/\left \vert \mathbf{\phi }\right \vert $ and
using the integral result $\int_{0}^{\infty }\frac{\rho d\rho }{\left(
1+\rho ^{2}\right) ^{2}}=\frac{1}{2}$, we end up with the following
expression%
\begin{equation}
\mathcal{J}^{i}\left( \mathbf{r}\right) =\frac{1}{2}\times \frac{1}{2\pi }%
\times \frac{4}{\left \vert \mathbf{\phi }\right \vert ^{2}}\varepsilon
^{ij}\phi _{a}\partial _{j}\phi _{b}\varepsilon ^{ab}
\end{equation}%
which is precisely Eq(\ref{cu}).

\begin{acknowledgement}
The authors would like to acknowledge "Acad\'{e}mie Hassan II des Sciences
et Techniques-Morocco" for financial support. L. B. Drissi acknowledges the
Alexander von Humboldt Foundation for financial support via the Georg
Forster Research Fellowship for experienced scientists (Ref 3.4 - MAR -
1202992).
\end{acknowledgement}

\end{document}